\documentclass[aps, prl, twocolumn,amsmath,amssymb,floatfix, reprint, groupedaddress, nobibnotes]{revtex4-2}

\usepackage[english]{babel}
\babeladjust{autoload.bcp47 = on}
\usepackage{graphicx}
\usepackage{mathtools}
\usepackage{bm}
\usepackage{braket}
\usepackage{microtype}
\usepackage[acronym]{glossaries}
\usepackage[hidelinks]{hyperref}
\usepackage{cleveref}
\usepackage{amsfonts}
\usepackage{xcolor}
\usepackage{physics}
\usepackage{times}
\hypersetup{colorlinks, allcolors=blue}

\usepackage{float}

\renewcommand{\section}[1]{\hspace*{1em}\textit{#1}---}
\newcommand{\shortbar}[1]{%
  \mkern1mu\overline{\mkern-1mu #1 \mkern-1mu}\mkern1mu}

\newcommand{\mr}[1]{\mathrm{#1}}

\newcommand{\figref}[2]{Fig.~\hyperref[#1]{\ref*{#1}\,(#2)}}

\usepackage{etoolbox} 
\pretocmd{\eqref}{Eq.~}{}{} 

\begin{document}

\title{Umklapp-enhanced high-$T_c$ topological superconductivity mediated by magnons in \textit{p}-wave magnets}

\author{Kristoffer Leraand}
 
\author{Asle Sudb{\o}}
\email{asle.sudbo@ntnu.no}
\affiliation{Center for Quantum Spintronics, Department of Physics, Norwegian University of Science and Technology, NO-7491 Trondheim, Norway}

\begin{abstract}
We show that magnon exchange can induce nodal topological superconductivity in a microscopic model of a \textit{p}-wave magnet. Solving the weak-coupling multiband gap equation for itinerant electrons exchange-coupled to localized spins, we find a nodal \textit{p}-wave gap and a critical temperature that rises sharply as the Fermi surface approaches the magnetic Brillouin-zone boundary. In this regime, small-reduced-momentum magnons can scatter electrons between Fermi-surface segments on opposite sides of the zone. These Umklapp processes enhance pairing because the contributions from different sites in the magnetic unit cell cancel less in their electron–magnon vertices than in those for ordinary scattering at comparable reduced momenta. Momentum-resolved winding numbers predict zero-energy Majorana flat bands at symmetry-preserving edges where the bulk nodes project to distinct momenta.
\end{abstract}
    
\maketitle

\section{Introduction}
Altermagnets have expanded the classification of collinear magnetic
order beyond conventional ferromagnetism and
antiferromagnetism~\cite{Smejkal-Sinova-Jungwirth, Agterberg24}.
Although their magnetic moments compensate to give zero net
magnetization, altermagnets exhibit even-parity spin-split electronic bands 
even without spin--orbit coupling.
This property has motivated studies of quasiparticle
dynamics~\cite{KL-SE}, collective excitations and
fluctuations~\cite{Liu2026, Ursin2026, Li2026, Lundemo2025,
Cichutek2025}, and
superconductivity~\cite{Leraand2025, MBS24, BBS23,
Madhusuthanan2026} in altermagnets.
Experimental evidence for altermagnetic
band splitting has been reported in
$\mr{MnTe}$~\cite{osumi2023MnTe, LeePRL2024MnTe, krempasky2024MnTe}
and $\mr{CrSb}$~\cite{reimers2023CrSb, Ding2024CrSb, Yang2025}.

Noncollinear magnetic order can also generate an odd-parity
electronic spin polarization through an isotropic exchange
interaction between itinerant electrons and localized
spins~\cite{Martin2012, Hayami2020a, Hayami2020b, Hayami2022,
Kudasov2024, Hellenes2024}.
In \textit{p}-wave magnets, the
electronic spin polarization changes sign under momentum
inversion~\cite{Hellenes2024, Brekke, Microscopic_origin_of_p-wave}.
Their magnetic textures preserve the composite symmetry
$\Theta=\mathcal{T}\boldsymbol{\tau}$, combining time reversal
$\mathcal{T}$ with a translation $\boldsymbol{\tau}$, typically
by half a magnetic unit cell. The corresponding spin texture
satisfies
$\textbf{S}_{\textbf{r}+\boldsymbol{\tau}}=-\textbf{S}_{\textbf{r}}$,
and the composite symmetry relates electronic states at opposite
momenta with opposite spin
polarizations~\cite{Brekke, Kudasov2024, Hellenes2024}.
This symmetry distinguishes \textit{p}-wave magnets from
more general helimagnets~\cite{Brekke, Hayami2022, Kudasov2024,
Chen2022, Mayo2025}.
Recent studies have investigated their microscopic
origin~\cite{Hubbard, Microscopic_origin_of_p-wave},
the effects of correlations and fluctuations in one
dimension~\cite{Sondre2026}, and electronic and magnetic
responses~\cite{Brekke, Edelstein, Maeland2026, Kagome_Smejkal}.

Several materials have been proposed as candidates for
\textit{p}-wave magnetism. 
One candidate particularly relevant here is
$\mr{CeNiAsO}$, in which the $\mr{Ce}$ spins are arranged
in coupled square layers with a coplanar noncollinear
spin structure~\cite{Hellenes2024, Wu2019, Maeland2026, Zhang2026}.
Other candidates include $\mr{Mn}_3\mr{GaN}$ and
$\mr{Ce}_3\mr{InN}$~\cite{Hellenes2024, Shi2016, Gabler2008}.
A coplanar spin helix has also been observed
in metallic
$\mr{Gd}_3(\mr{Ru}_{1-\delta}\mr{Rh}_{\delta})_4\mr{Al}_{12}$
near $\delta=0.05$~\cite{Nature}.
These developments motivate investigating magnon-mediated
pairing in a generic square-lattice model with coplanar
\textit{p}-wave magnetic order.

The symmetry of $p$-wave magnets permits conventional zero-momentum superconductivity even in the
presence of a large magnetic exchange splitting~\cite{p-wave_SC, Smejkal_SC}.
The competition and coexistence of different pairing symmetries
have been investigated~\cite{Brekke_anb_SC}.
Studies using phenomenological attractive interactions have
demonstrated that \textit{p}-wave magnets can support nodal
topological superconducting states~\cite{KimV2}, and
related work has extended this result to other odd-parity magnetic
systems~\cite{Hidden_zeemanV2, X-wave}. Bulk-boundary correspondence allows
topological superconductors to support Majorana boundary
modes at their edges~\cite{Review_TSC, Qi2011, Bernevig2013, Leijnse2012,
Zlotnikov2021}.
In nodal superconductors, momentum-resolved winding numbers can protect zero-energy Majorana flat
bands at suitable boundaries~\cite{Nodal_SC_review, KimV2, X-wave}.
Microscopic studies have also predicted magnon-mediated
topological superconductivity at interfaces between normal metals and either
skyrmion crystals~\cite{KM_SC} or coplanar helical
magnets~\cite{Sara}.
These studies assume spin-degenerate normal-metal states, leaving open the question of how pairing
operates in the odd-parity spin-split bands of a $p$-wave magnet.
We therefore ask whether magnon exchange
can generate nodal topological superconductivity in a
microscopic model that retains both the spin-split electron
bands and the magnetic excitations of a \textit{p}-wave magnet.

In this Letter, we show that magnon exchange can mediate a nodal
topological superconducting state in a microscopic model of a
\textit{p}-wave magnet. We consider itinerant electrons
exchange-coupled to localized spins on a square lattice and derive
the effective electron-electron interaction from the
electron-magnon coupling. Solving the resulting multiband gap
equation within a weak-coupling approach, we obtain a superconducting
gap with a \textit{p}-wave momentum structure and a critical
temperature that depends strongly on the Fermi-surface geometry.
We characterize the superconducting topology by calculating momentum-resolved winding numbers, which predict Majorana flat bands at edges onto which
the bulk nodes project at distinct momenta.

\section{Spin Model}
We model the p-wave magnet using localized spins on a square lattice and allow the on-site spins to interact with their nearest neighbors. The ferromagnetic exchange has different interaction strengths in the \textit{x} ($J_x$) and \textit{y} ($J_y$) directions. We include a Dzyaloshinskii--Moriya interaction (DMI), which favors perpendicular alignment of neighboring spins. Competition between the DMI and the ferromagnetic exchange sets the period of the spin texture. We also include an easy-axis anisotropy along the \textit{z}-axis. The spin Hamiltonian is then given by
\begin{equation}
    \begin{aligned}
        H_\mr{s} = &-J_x\sum_{\langle i,j\rangle_x}\textbf{S}_{i}\cdot \textbf{S}_{j} -J_y\sum_{\langle i,j\rangle_y}\textbf{S}_{i}\cdot \textbf{S}_{j}
        \\
        &+\sum_{\langle i,j\rangle}\textbf{D}_{ij}\cdot\left(\textbf{S}_i\cross\textbf{S}_j\right)
        -K_z\sum_i \left(S^z_i\right)^2,
    \end{aligned}
\end{equation}
where $\textbf{S}_i$ is the spin at site \textit{i}, and $\langle i,j\rangle_{x(y)}$ denotes nearest-neighbor bonds in the \textit{x} (\textit{y})-direction. The DMI vector is bond-dependent and given by $\textbf{D}_{ij}=D\hat{\textbf{r}}_{ij}\cross\hat{\textbf{z}}$, where $\hat{\textbf{r}}_{ij}$ is a unit vector pointing from site \textit{i} to site \textit{j}. $K_z$ is the easy-axis anisotropy strength. We take all $J_x$, $J_y$, $D$, and $K_z$ to be positive. 

The ground state in \figref{main:fig:1}{a} is obtained by minimizing the classical spin energy, as described in Sec.~S.II in the Supplemental Material (SM)~\cite{SM}. 
The Supplemental Material also describes how we obtain the magnon spectrum using the
Holstein–Primakoff transformation.
The magnon energy bands are shown in  \figref{main:fig:1}{b}.
\begin{figure}[tb]
    \centering
    \includegraphics[width=\linewidth]{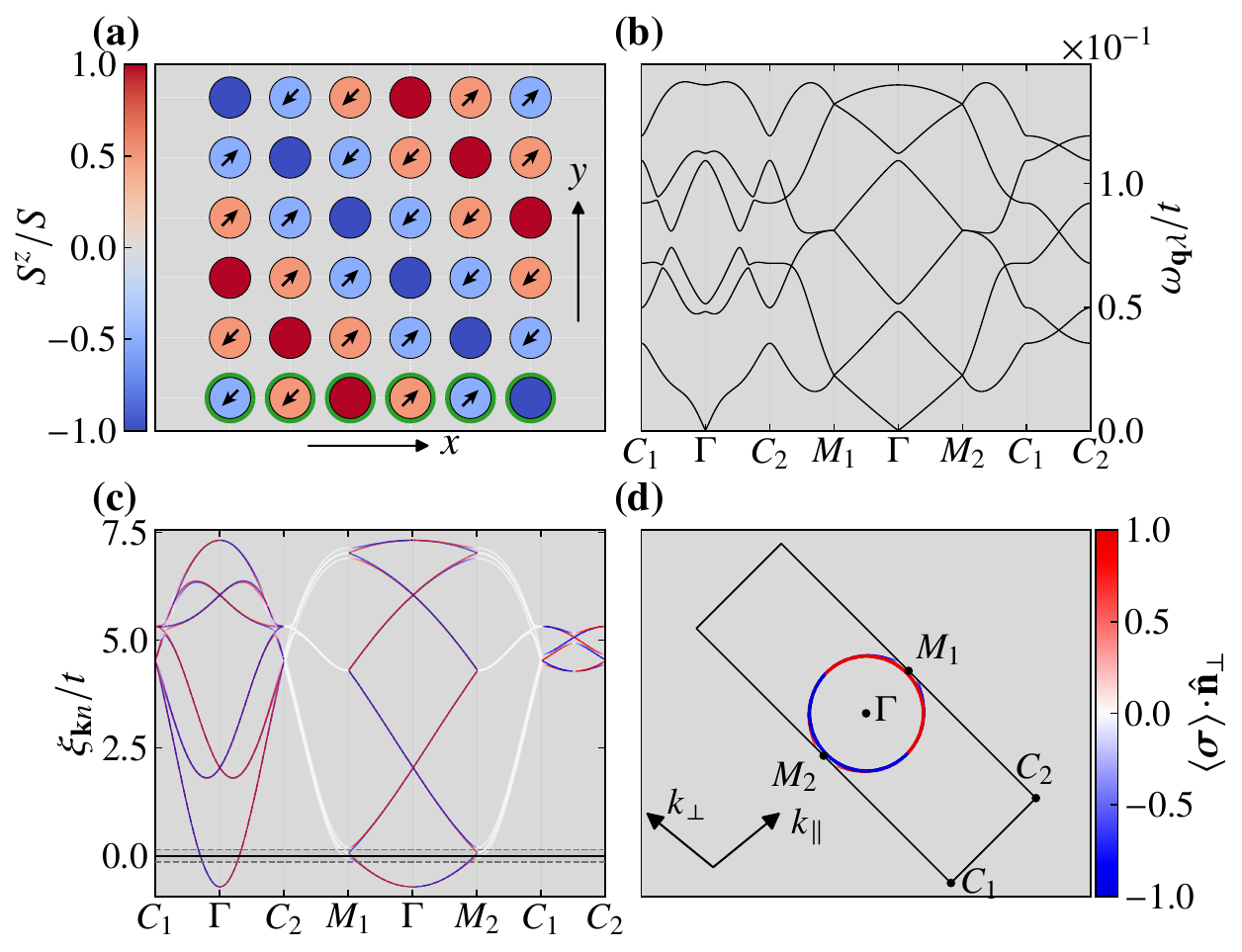}
    \caption{
    \textbf{(a)} Classical magnetic ground state. The arrows show the in-plane spin components $(S_x,S_y)/S$, and the colors show the out-of-plane component $S_z/S$. Green circles mark the six-site magnetic unit cell with lattice vectors $\textbf{T}_1=a\left(6,0,0\right)$ and $\textbf{T}_2=a\left(-1,1,0\right)$. 
    \textbf{(b)} Magnon bands along the indicated path through the magnetic Brillouin zone.
    \textbf{(c)} Electron bands along the same path. The solid black line marks the Fermi level, and the dashed gray lines mark the pairing-shell limits $\xi=\pm\omega_\mr{M}$, where $\omega_\mr{M}$ is the maximum magnon energy.
    \textbf{(d)} Fermi surface in the magnetic Brillouin zone, with the symmetry points and the $k_\parallel$ and $k_\perp$ directions indicated. The colors in \textbf{(c)} and \textbf{(d)} indicate the electronic spin projection perpendicular to the plane of the ordered spin moments. The model parameters are given in Table SI in the SM~\cite{SM}.
    }
    \label{main:fig:1}
\end{figure}

\section{Electron Model}
We introduce itinerant electrons on the lattice in \figref{main:fig:1}{a}, with nearest- and next-nearest-neighbor hopping
amplitudes $t$ and $t_2$, respectively.
The chemical potential $\mu$ controls the electron filling, and an on-site exchange couples the electron spins to the local moments. The electronic 
Hamiltonian is given by
\begin{equation}
\begin{aligned}
\label{main:eq:Electron Hamiltonian}
    H_\mr{el}=&-t\sum_{\langle i,j\rangle\sigma}c_{i\sigma}^\dag c_{j\sigma}-t_2\sum_{\langle\langle i,j\rangle\rangle\sigma}c_{i\sigma}^\dag c_{j\sigma}
    \\
    &-\mu\sum_{i\sigma}c_{i\sigma}^\dag c_{i\sigma}-J_\mr{sd}\sum_{i\sigma\sigma'}\textbf{S}_i\cdot c_{i\sigma}^\dag \boldsymbol{\sigma}_{\sigma\sigma'}c_{i\sigma'},
\end{aligned}
\end{equation}
where $c_{i\sigma}^{\left(\dag\right)}$ annihilates (creates) an electron at site $i$ with spin $\sigma$. We denote the set of nearest-neighbor bonds by $\langle i,j\rangle$, and $\langle\langle i,j\rangle\rangle$ is the set of next-nearest-neighbor bonds.
$J_{\mathrm{sd}}$ sets the exchange strength between the itinerant electrons and the localized spins. Here, $\boldsymbol{\sigma}$ is a vector of Pauli matrices. 

We Fourier transform the Hamiltonian as described in Sec.~S.IV of the SM~\cite{SM}, and diagonalize momentum-space Hamiltonian by a unitary transformation, $\textbf{c}_\textbf{k}=U(\textbf{k})\textbf{d}_\textbf{k}$. Here, $U(\textbf{k})$ is a unitary matrix and the \textit{d} operators describe long-lived fermionic quasiparticles,
\begin{equation}
    H_\mr{el} = \sum_{\textbf{k}n}\xi_{\textbf{k}n}d_{\textbf{k}n}^\dag d_{\textbf{k}n}.
\end{equation}
Here, \textit{n} is a band label and $\xi_{\textbf{k}n}$ is the fermionic energy with momentum $\textbf{k}$ in band \textit{n}. The electron bands are shown in \figref{main:fig:1}{c}, and the Fermi surface (FS) is shown in \figref{main:fig:1}{d}. The defining symmetry of \textit{p}-wave magnets, $\Theta=\mathcal{T}\boldsymbol{\tau}$, require the electronic states to satisfy $\Theta \xi_{\textbf{k},\sigma}\Theta^{-1} = \xi_{-\textbf{k},-\sigma}$~\cite{Brekke}, as illustrated in \figref{main:fig:1}{c} and \figref{main:fig:1}{d}.
The spin quantization axis for a coplanar spin texture is perpendicular to the spin plane ~\cite{Spin_OOP}.

\section{Interaction}
\begin{figure}[tb]
    \centering
    \includegraphics[width=\linewidth]{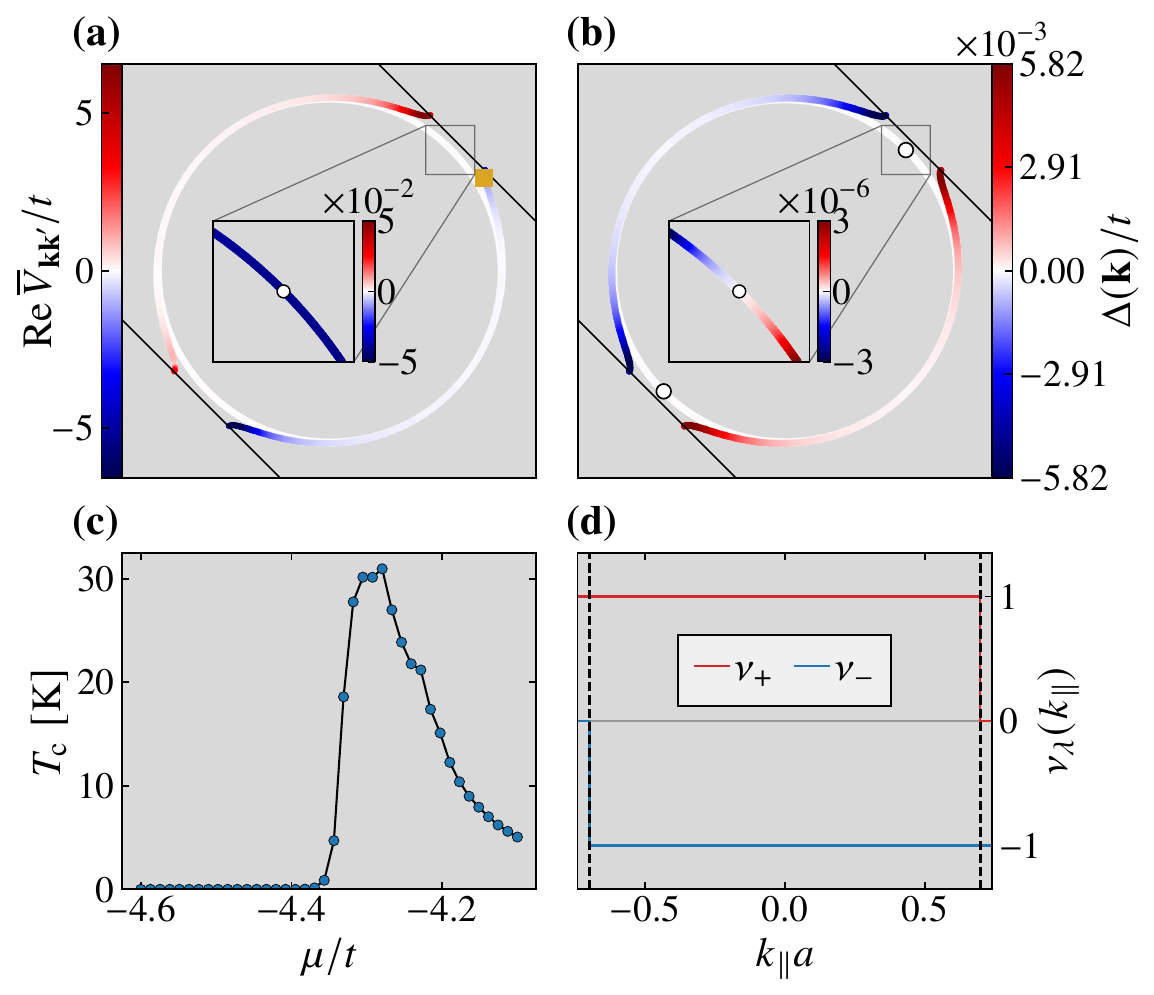}
    \caption{\textbf{(a)} The real part of the effective electron--electron interaction as a function of $\textbf{k}'$, where \textbf{k} is marked by the yellow square. The inset enlarges the region around a gap node, marked by a white circle. \textbf{(b)} Superconducting gap at zero temperature. White circles mark the gap nodes, and the inset enlarges the
region around one node.\textbf{(c)}  Critical temperature as a function of chemical potential. \textbf{(d)} One-dimensional winding number in each sector as a function of momentum parallel to the edge. The black dashed lines indicate the momenta corresponding to the gap nodes. The model parameters are given in Table SI in the SM~\cite{SM}.}
    \label{main:fig:2}
\end{figure}

The on-site exchange in \eqref{main:eq:Electron Hamiltonian} produces the spin-split electron bands through the static magnetic texture and couples the electrons to magnons through fluctuations about that texture. To derive the electron-magnon coupling, we expand the localized spins about their classical orientations using the Holstein--Primakoff transformation.
Keeping the terms linear in the boson operators and expressing the interaction in terms of the electron bands and magnon modes gives
\begin{equation}
\label{main:eq:electron magnon band coupling}
\begin{aligned}
    H_\mr{em}
    =\sum_{\textbf{k}\textbf{q}}\sum_{\lambda nn'}
    \Big[
    &g_{\lambda nn'}^{(1)}(\textbf{k}+\textbf{q},\textbf{k})
    \alpha_{-\textbf{q},\lambda}^\dag
    \\
    +\,
    &g_{\lambda nn'}^{(2)}(\textbf{k}+\textbf{q},\textbf{k})
    \alpha_{\textbf{q}\lambda}
    \Big]
    d_{\textbf{k}+\textbf{q},n}^\dag d_{\textbf{k}n'}.
\end{aligned}
\end{equation}
The vertices $g^{(1)}$ and $g^{(2)}$ describe an electron scattering from $(\textbf{k},n')$ to $(\textbf{k}+\textbf{q},n)$ by emitting a magnon with momentum $-\textbf{q}$ or absorbing one with momentum $\textbf{q}$, respectively. Explicit expressions for the vertices are given in Sec.~S.V in the SM~\cite{SM}. Next, we apply a Schrieffer--Wolff transformation to obtain an effective electron--electron interaction. To second order in the electron--magnon coupling, one electron emits a virtual magnon, and the other absorbs it. We restrict the interaction to Cooper pairs with zero total momentum, so it scatters a pair from $(\textbf{k}',-\textbf{k}')$ to $(\textbf{k},-\textbf{k})$. The resulting effective Hamiltonian is given by
\begin{equation}
\label{main:eq:effective elec-elec}
    H_{\mr{eff}}=\sum_{\textbf{k}\textbf{k}'}\sum_{nm,n'm'}V_{\textbf{k}\textbf{k}'}^{nm,n'm'}d_{\textbf{k}n}^\dag d_{-\textbf{k},m}^\dag d_{-\textbf{k}',m'}d_{\textbf{k}',n'},
\end{equation}
where
\begin{widetext}
\begin{equation}
\label{main:eq:SW interaction}
\begin{aligned}
    V_{\textbf{k}\textbf{k}'}^{nm,n'm'}
    = -\frac{1}{4}\sum_\lambda \Bigg[
    &\frac{2\omega_{\textbf{k}'-\textbf{k},\lambda}+\delta\xi_+-\delta\xi_-}
    {\bigl(\omega_{\textbf{k}'-\textbf{k},\lambda}+\delta\xi_+\bigr)
     \bigl(\omega_{\textbf{k}'-\textbf{k},\lambda}-\delta\xi_-\bigr)}
    \,g_{\lambda nn'}^{(1)}(\textbf{k},\textbf{k}')
    g_{\lambda mm'}^{(2)}(-\textbf{k},-\textbf{k}') \\
    +&
    \frac{2\omega_{\textbf{k}-\textbf{k}',\lambda}+\delta\xi_--\delta\xi_+}
    {\bigl(\omega_{\textbf{k}-\textbf{k}',\lambda}+\delta\xi_-\bigr)
     \bigl(\omega_{\textbf{k}-\textbf{k}',\lambda}-\delta\xi_+\bigr)}
    \,g_{\lambda nn'}^{(2)}(\textbf{k},\textbf{k}')
    g_{\lambda mm'}^{(1)}(-\textbf{k},-\textbf{k}')
    \Bigg].
\end{aligned}
\end{equation}
\end{widetext}
Here $\delta\xi_+ = \xi_{\textbf{k}n}-\xi_{\textbf{k}'n'}$ and
$\delta\xi_- = \xi_{-\textbf{k},m}-\xi_{-\textbf{k}',m'}$ are the changes in the energies of the two electrons from the incoming to the outgoing pair.
We evaluate the interaction on the FS, where the electronic energy differences $\delta\xi_+$ and $\delta\xi_-$ vanish.
Each product of vertices corresponds to one of the two possible emitters of the virtual magnon. The electrons acquire opposite momentum transfers, $\textbf{q}=\textbf{k}-\textbf{k}'$, and $-\textbf{q}$. Exchanging the two electrons in either the incoming or the outgoing pair reverses the sign of the corresponding fermion operator product. We therefore combine the four orderings of the incoming and outgoing states to obtain the antisymmetrized
interaction
\begin{equation}
\label{main:eq:Pauli antisymmetrized interaction}
\begin{aligned}
    \shortbar{V}_{\textbf{k}\textbf{k}'}^{nm,n'm'}=\frac12\Big(
    &V_{\textbf{k}\textbf{k}'}^{nm,n'm'}
    +V_{-\textbf{k},-\textbf{k}'}^{mn,m'n'}
    \\
    -&V_{-\textbf{k},\textbf{k}'}^{mn,n'm'}
    -V_{\textbf{k},-\textbf{k}'}^{nm,m'n'}\Big).
\end{aligned}
\end{equation}
\figref{main:fig:2}{a} shows the real part of $\shortbar{V}$ for the dominant band combination as a function of $\textbf{k}'$ along the FS.

\section{Superconductivity}
To study superconductivity, we combine the normal-state band Hamiltonian with the antisymmetrized pairing interaction to obtain
\begin{equation}
\begin{aligned}
    H = &\sum_{\textbf{k}n}\xi_{\textbf{k}n}d_{\textbf{k}n}^\dag d_{\textbf{k}n}
    \\
    +\frac12&\sum_{\textbf{k}\textbf{k}'}\sum_{nm,n'm'}\shortbar{V}^{nm,n'm'}_{\textbf{k}\textbf{k}'} d_{\textbf{k}n}^\dag d_{-\textbf{k},m}^\dag d_{-\textbf{k}',m'}d_{\textbf{k}',n'}.
\end{aligned}
\end{equation}
In the mean-field approximation, we neglect terms quadratic in the fluctuations of the Cooper-pair operators
about their mean values. We use a multiband BCS description that allows an arbitrary number of Fermi-surface states at each pairing
momentum.
 The resulting Bogoliubov-de Gennes (BdG) Hamiltonian contains the gap matrix $\Delta_{\textbf{k}nm}$, whose elements describe pairing between the states $(\textbf{k},n)$ and $(-\textbf{k},m)$.
Minimizing the mean-field grand potential with respect to the gap matrix gives the self-consistency equation for the gap, as derived in Sec.~S.VI of the SM~\cite{SM}. For a fixed outgoing pair $(\textbf{k},n)$, $(-\textbf{k},m)$, the gap equation sums over the amplitudes of incoming pairs $(\textbf{k}',n')$, $(-\textbf{k}',m')$, weighted by the interaction that scatters each incoming pair into the outgoing pair. We restrict this sum to pairs for which both electron energies lie within $\omega_\mr{M}$ of the Fermi level, where $\omega_\mr{M}$ is the maximum magnon energy. Within this shell, we approximate the gap and the interaction to be constant along the direction normal to the Fermi surface.
We retain the momentum dependence of the electronic dispersions throughout the shell. In the continuum limit, the sum over $\textbf{k}'$ becomes an integral over the pairing shell.
For each band $n'$, we use the band energy $\xi=\xi_{\textbf{k}'n'}$ and the arc length s along the corresponding constant-energy contour $C_{n'}(\xi)$ as integration variables. In these variables, the gap equation becomes
\begin{equation}
\label{main:eq:Gap FS}
    \begin{aligned}
        &\Delta_{\textbf{k}_\mr{F}nm} = -\frac{N_\mr{c}}{A_\mr{BZ}}\sum_{n'm'}\int_{-\omega_\mr{M}}^{\omega_\mr{M}}\!\!\dd\xi\oint_{C_{n'}(\xi)}\frac{\dd s}{\left|\nabla_{\textbf{k}'}\xi_{\textbf{k}'n'}\right|}
        \\[0.5ex]
        &\times\shortbar{V}^{nm,n'm'}_{\textbf{k}_\mr{F}\textbf{k}_\mr{F}'}\left\langle d_{-\textbf{k}',m'}d_{\textbf{k}'n'}\right\rangle\Theta\left(\omega_\mr{M}-|\xi_{-\textbf{k}',m'}|\right),
    \end{aligned}
\end{equation}
where $N_\mr{c}$ is the number of unit cells and $A_\mr{BZ}$ is the area of the first BZ. The outgoing momentum $\textbf{k}_\mr{F}$ lies on the FS for band \textit{n}, and $\textbf{k}'_\mr{F}$ is the projection of $\textbf{k}'$ onto the FS of band $n'$ along the local normal. The limits on $\xi$ enforce the energy cutoff for the first electron $(\textbf{k}',n')$, and the Heaviside function $\Theta$ applies the same cutoff to its partner $(-\textbf{k}',m')$.

We solve the gap equation at the critical temperature and at zero temperature. At the critical temperature,
we expand the pair amplitude in \eqref{main:eq:Gap FS} in powers of the infinitesimal gap 
\begin{align}
\label{main:eq:Linear expectation value}
    \left\langle d_{-\textbf{k},m}d_{\textbf{k}n}\right\rangle &=\chi_{\textbf{k}nm}(T)\Delta_{\textbf{k}_\mr{F}nm}+\mathcal{O}\left(|\Delta|^2\right),
    \\[0.5ex]
    \chi_{\textbf{k}nm}(T) &= \frac{1-f(\xi_{\textbf{k}n})-f(\xi_{-\textbf{k},m})}{\xi_{\textbf{k}n}+\xi_{-\textbf{k},m}},
\end{align}
where $f$ is the Fermi-Dirac distribution. Inserting this expression into the gap equation yields an eigenvalue problem, $\Delta=L(T)\Delta$, where $L(T)$ is the linearized pairing kernel obtained from \eqref{main:eq:Gap FS}. The critical temperature $T_\mr{c}$ is the highest temperature at which the leading eigenvalue of $L(T)$ equals one, marking the transition from the normal state to the superconducting state. The corresponding eigenvector gives the gap profile at the transition up to an overall normalization.

At zero temperature, the pair amplitude in \eqref{main:eq:Gap FS} becomes
\begin{equation}
    \left\langle d_{-\textbf{k},m} d_{\textbf{k}n}\right\rangle_{T=0} = \!\!\!\!\!\sum_{\nu=1}^{N_{\textbf{k}}+N_{-\textbf{k}}}\!\!\big[W_{\textbf{k}}\big]_{n\nu} \big[W^*_{\textbf{k}}\big]_{N_\textbf{k}+m,\nu}\Theta(-E_{\textbf{k}\nu}),
\end{equation}
where $N_\textbf{k}$ is the number of states within the pairing shell at momentum $\textbf{k}$. $\nu$ labels the different BdG bands, while $W_\textbf{k}$ and $E_\textbf{k}$ are matrices. The columns of $W_\textbf{k}$ are eigenvectors of the BdG Hamiltonian and $E_\textbf{k}$ is a diagonal matrix with the corresponding eigenvalues, $\mathcal{H}_\mr{BdG}(\textbf{k})W_\textbf{k}=W_\textbf{k}E_\textbf{k}$.
The zero-temperature gap equation is nonlinear because the pair amplitude depends on the gap through the BdG eigenvalues and eigenvectors. We solve it self-consistently, using the gap profile obtained at $T_\mr{c}$ as the initial guess.
The zero-temperature gap exhibits a nodal \textit{p}-wave structure as shown in \figref{main:fig:2}{b}.
Comparing the interaction in \figref{main:fig:2}{a} with the gap, we can see that the gap and the interaction have opposite signs. Since the interaction in the gap equation enters with a negative sign, this sign mismatch reinforces the superconducting gap. 

The critical temperature shown in \figref{main:fig:2}{c} varies strongly with the chemical potential and peaks at about $30$~K in our weak-coupling regime. It rises sharply as the Fermi surface approaches the Brillouin-zone boundary. In this case, Umklapp scattering connects states on opposite sides of the zone through small reduced momentum transfers.
Removing these short Umklapp transfers substantially reduces the leading pairing eigenvalue near the peak. At the peak, the electron-magnon amplitudes from different magnetic sites cancel less for Umklapp processes than for ordinary processes at comparable reduced transfers. At higher filling, the FS still reaches the zone boundary, but the contour starts to bend away from the boundary. Because of this, the positive pairing contributions from short Umklapp transfers decrease, accounting for the fall in $T_c$.
Umklapp scattering has also been found to play an important role in magnon-mediated superconductivity in related systems with a Kondo coupling~\cite{Sara, Thingstad-Erlandsen}. We analyze how these Umklapp processes affect the pairing kernel in Sec.~S.VII of the SM~\cite{SM}.

No superconducting phase transition has been observed in $\mr{CeNiAsO}$ down to $30$~mK at ambient pressure~\cite{Luo2011CeNiAsO, Luo2014QuantumCriticality, Li2014Kramers}. 
Mapping the normal-state Fermi surface relative to this boundary could guide chemical-potential tuning, for example by electrostatic gating, toward fillings favorable for superconductivity.
Fillings where a large portion of the FS lies near the magnetic BZ boundary will favor pairing through short Umklapp transfers.

\section{Topology}
The superconducting gap in \figref{main:fig:2}{b} has a nodal p-wave structure. The bulk nodes prevent us from using
the global invariants employed for fully gapped superconductors~\cite{KM_SC, Sara, Review_TSC}. Instead, we characterize the nodal superconducting state using a one-dimensional winding number for each conserved momentum $k_\parallel$, parallel to the edge of our open system~\cite{KimV2, Review_TSC, Top1, Top2, Top3, Top4, Nodal_SC_review}.
This invariant is well defined whenever the BdG spectrum is gapped along the corresponding momentum cut. Each node can carry a topological charge, acting as a source for the topological invariant~\cite{Review_TSC, Nodal_SC_review}. Fig.~S3 of the SM~\cite{SM}, shows that the nodes carry opposite topological charges. The two bulk nodes lie along the line $k_y=k_x$, as seen in \figref{main:fig:2}{b}, so we choose an edge parallel to $(1,1)$. In this case $k_\parallel\propto k_x+k_y$, so the nodes project onto distinct momenta and the winding number is nonzero. The two momentum directions $k_\parallel$ and $k_\perp$ are marked in \figref{main:fig:1}{d}. We define the one-dimensional topological invariant as
\begin{equation}
\label{main:eq:sector winding}
    \nu_\lambda(k_\parallel)
    =
    \frac{1}{2\pi i}
    \oint_{\mathcal{C}_\lambda}
    \!\!\!\dd k_\perp\,
    \partial_{k_\perp}
    \ln\det q_\lambda(k_\parallel,k_\perp),
\end{equation}
where $\mathcal{C}_\lambda$ is a closed transverse momentum path at fixed $k_\parallel$. The sector index $\lambda=\pm1$ labels the blocks of the bulk BdG Hamiltonian under the symmetry $U_\tau$. This symmetry combines a spin rotation by $\pi$ about the normal to the spin plane with a translation by half a magnetic unit cell. The matrix $q_\lambda$ is the off-diagonal block of the BdG Hamiltonian in the chiral basis for sector $\lambda$. For a single-band superconductor, we have $q_\lambda(\textbf{k})=\xi_{\textbf{k}\lambda}+i\Delta_{\textbf{k}\lambda}$. The invariant measures how many times this complex number winds around the origin. In the multiband
case, it measures the winding of $\mr{det}q_\lambda$. We construct the chiral basis and calculate the winding numbers in each sector in Secs.~S.VIII~A and S.VIII~B of the SM~\cite{SM}. 
The sector winding numbers are nonzero on the gapped cuts between the projected bulk nodes shown in \figref{main:fig:2}{d}.
By bulk-boundary correspondence, a nonzero winding over a finite interval of $k_\parallel$ predicts a zero-energy flat band at an edge that preserves the symmetries protecting the winding number. The BdG particle-hole symmetry maps a state at $(E,k_\parallel)$ to a partner at $(-E,-k_\parallel)$. At zero energy, the partner is also a zero mode, and its quasiparticle operator is the adjoint of the quasiparticle operator for the original state, $\gamma_{-k_\parallel}=\gamma_{k_\parallel}^\dag$. Hence, the combination $\gamma_{k_\parallel} + \gamma_{-k_\parallel}$ is self-adjoint, and the zero-energy flat bands are Majorana flat bands.

\section{Conclusion}
We have shown that magnons can mediate nodal topological superconductivity in a microscopic model of a \textit{p}-wave magnet. Solving the resulting multiband gap equation within a weak-coupling approach yields a superconducting gap with a \textit{p}-wave momentum structure and a critical temperature that peaks at about $30$~K. 
The critical temperature rises sharply as tuning the chemical potential brings the Fermi surface close to the Brillouin-zone boundary, where Umklapp scattering with small reduced momentum transfers enhances pairing. 
The momentum-resolved winding numbers are nonzero in both sectors, predicting zero-energy Majorana flat bands at suitably oriented edges that preserve the symmetries that protect the winding number. Our results establish a microscopic route to topological superconductivity in which the magnetic order shapes the electronic spin texture and magnetic fluctuations provide the pairing
interaction. As in the nodal \textit{d}-wave cuprates, a high pairing scale may help nodal superconductivity survive weak disorder by keeping the coherence length short compared with the elastic mean free path.

\begin{acknowledgments}
We thank Hans Gløckner Giil, Sondre Duna Lundemo, and Bjørnulf Brekke for helpful discussions. KL and AS were supported by the Research Council of Norway (RCN) through its Centres of Excellence funding scheme, Project No.\ 262633, ``QuSpin'', RCN Project No.\ 323766, as well as COST Action CA21144  ``Superconducting Nanodevices and Quantum Materials for Coherent Manipulation". 
\end{acknowledgments}

\bibliography{Refs.bib, manual_refs.bib}
\end{document}


\title{Supplemental Material \\ Umklapp-enhanced high-$T_c$ topological superconductivity mediated by magnons in \textit{p}-wave magnets}

\author{Kristoffer Leraand}
\affiliation{Center for Quantum Spintronics, Department of Physics, Norwegian University of Science and Technology, NO-7491 Trondheim, Norway}

\author{Asle Sudb\o}
\affiliation{Center for Quantum Spintronics, Department of Physics, Norwegian University of Science and Technology, NO-7491 Trondheim, Norway}

\maketitle

\onecolumngrid

\setcounter{tocdepth}{2}
\tableofcontents
\clearpage

\section{Introduction}
This Supplemental Material presents the derivations and numerical details underlying the results of the Letter. In Section~\ref{sm:sec:Gnd}, we determine the classical magnetic ground state, which serves as the starting point for the Holstein--Primakoff expansion and the derivation of the magnon spectrum in Section~\ref{sm:sec:HP}. After diagonalizing the normal-state electron Hamiltonian in Section~\ref{sm:sec:Elec}, we derive the electron--magnon vertices and the effective pairing interaction in Section~\ref{sm:sec:EL-mag}.  
In Section~\ref{sm:sec:SC}, we derive the multiband gap equation. We linearize the equation to determine the critical temperature and use its full nonlinear form to obtain the zero-temperature gap.
Section~\ref{sm:sec:Tc} examines the role of Fermi-surface geometry and Umklapp scattering in the rise and fall of the critical temperature as the chemical potential varies. In Section~\ref{sm:sec:Topology}, we analyze the symmetries of fixed-momentum cuts, derive the sector-resolved winding numbers, and discuss the bulk nodes and their edge projections. Section~\ref{sm:sec:Numerics} explains how we relate the magnon Bogoliubov transformation at opposite momenta and numerically solve the gap equations. Section~\ref{sm:sec:Numerical params} lists the parameter values used in the calculations.

\section{Classical ground state}
\label{sm:sec:Gnd}
For the spin system, we consider a Hamiltonian with ferromagnetic Heisenberg exchange between nearest neighbors, with strengths $J_x$ and $J_y$ along the \textit{x} and \textit{y} directions, respectively. The competition between the exchange and the bond-dependent Dzyaloshinskii-Moriya interaction (DMI) favors a helical spin texture. We also include a weak easy-axis anisotropy $K_z$ along \textit{z}, which opens a small gap in the magnon spectrum.
This Hamiltonian is given by
\begin{equation}
    \begin{aligned}
    \label{sm:eq:Spin Hamiltonian}
        H_\mr{s} =  -\sum_{\langle i,j\rangle}J_{ij}\textbf{S}_{i}\cdot \textbf{S}_{j}
        &+\sum_{\langle i,j\rangle}\textbf{D}_{ij}\cdot\left(\textbf{S}_i\cross\textbf{S}_j\right)
        -K_z\sum_i \left(S^z_i\right)^2,
    \end{aligned}
\end{equation}
where summing over $\langle i,j\rangle$ means to sum over each nearest-neighbor bond once and $J_{ij}=J_{x(y)}$ for neighbors along the \textit{x} (\textit{y}) direction. The bond-dependent DMI vector is given by $\textbf{D}_{ij}=D\hat{\textbf{r}}_{ij}\cross\hat{\textbf{z}}$, where $\hat{\textbf{r}}_{ij}$ is a unit vector pointing from site \textit{i} to site \textit{j}.
To solve for the classical ground state, we set $\textbf{S}_i=S\hat{\textbf{n}}_i$, where $S$ is the fixed spin length and $\hat{\textbf{n}}_i$ is a unit vector giving the spin direction at site \textit{i}. We parametrize the unit vector as
\begin{equation}
    \hat{\textbf{n}}_i=\begin{pmatrix}
        \sin(\theta_i)\cos(\varphi_i)\\[0.5ex]
        \sin(\theta_i)\sin(\varphi_i)\\[0.5ex]
        \cos(\theta_i)
    \end{pmatrix}.
\end{equation}
Substituting $\textbf{S}_i=S\hat{\textbf{n}}_i$ into the spin Hamiltonian, \eqref{sm:eq:Spin Hamiltonian}, the classical energy per site is given by
\begin{equation}
    E_\mr{cl}=\frac {S^2}{N}\left[
        -\sum_{\langle i,j\rangle}
        J_{ij}\hat{\textbf{n}}_i\cdot\hat{\textbf{n}}_j
        +\sum_{\langle i,j\rangle}
        \textbf{D}_{ij}\cdot
        \left(\hat{\textbf{n}}_i\cross\hat{\textbf{n}}_j\right)
        -K_z\sum_i\left(n_i^z\right)^2
        \right],
\end{equation}
where \textit{N} is the total number of magnetic sites.
The classical ground state is given by the set of angles that minimizes this energy, so a stationary point satisfies
\begin{equation}
    \frac{\partial E_\mr{cl}}{\partial\theta_i}=0,
    \qquad
    \frac{\partial E_\mr{cl}}{\partial\varphi_i}=0,
\end{equation}
for all sites.
Let $q_i$ denote either $\theta_i$ or $\varphi_i$, then the stationary condition for the classical spin energy is given by
\begin{equation}
    \begin{aligned}
        \frac{\partial E_\mr{cl}}{\partial q_i} = \frac{S^2}{N}\left[-\sum_{j}J_{ij}\frac{\partial\hat{\textbf{n}}_i}{\partial q_i}\cdot\hat{\textbf{n}}_j+\sum_j\textbf{D}_{ij}\cdot\left(\frac{\partial\hat{\textbf{n}}_i}{\partial q_i}\cross\hat{\textbf{n}}_j\right)-2K_zn_i^z\frac{\partial n_i^z}{\partial q_i}
        \right]=0,
    \end{aligned}
\end{equation}
where summing over \textit{j} means to sum over all nearest neighbors of site \textit{i}.
Numerically, when solving for the set of angles that minimizes the classical energy, we test several initial spin configurations. These include helical states in different planes and random spin configurations. For each configuration, we minimize the classical energy with the L-BFGS-B algorithm~\cite{L-BFGS-B}, using the analytical expressions for the derivatives. It is helpful to define the local effective field by
\begin{equation}
    \textbf{h}_i=-\frac{N}{S^2}\pdv{E_\mr{cl}}{\hat{\textbf{n}}_i}= \sum_j\left(J_{ij}\hat{\textbf{n}}_j+\textbf{D}_{ij}\cross\hat{\textbf{n}}_j\right)+2K_zn_i^z\hat{\textbf{z}}.
\end{equation}
An infinitesimal change in the spin direction at site \textit{i} gives the first-order energy change
\begin{equation}
    \delta E_\mr{cl}=-\frac{S^2}{N}\textbf{h}_i\cdot\delta\hat{\textbf{n}}_i.
\end{equation}
Since $\hat{\textbf{n}}_i$ is a unit vector with a fixed length, any variation along the direction of $\hat{\textbf{n}}_i$ will vanish, $\hat{\textbf{n}}_i\cdot\delta\hat{\textbf{n}}_i=0$.
Only the component of $\textbf{h}_i$ perpendicular to $\hat{\textbf{n}}_i$ contributes to the energy change. At a stationary point, the first-order energy change vanishes, so must this component of $\textbf{h}_i$. The stationary conditions is therefore
\begin{equation}
    \hat{\textbf{n}}_i\cross \textbf{h}_i=0,
\end{equation}
for every site. 
Among the converged configurations, we select the one with the lowest energy and verify that $ \hat{\textbf{n}}_i\cross \textbf{h}_i$ vanishes within numerical tolerance at every site.

\section{Magnon Hamiltonian}
\label{sm:sec:HP}
Magnons describe small fluctuations about the classical spin configuration found in Section~\ref{sm:sec:Gnd}. We expand the spin Hamiltonian to quadratic order in Holstein--Primakoff bosons and diagonalize it to obtain the magnon energies and the Bogoliubov transformation. Both enter the effective pairing interaction derived in Section~\ref{sm:sec:EL-mag}. Because the classical spin direction varies from site to site, we introduce a local orthonormal basis at each site \textit{i}, with the local \textit{z} axis along $\hat{\textbf{n}}_i$. In terms of $\theta_i$ and $\varphi_i$, the basis vectors are given by
\begin{equation}
\label{sm:eq:local spin basis}
    \hat{\textbf{e}}_i^1 = 
    \begin{pmatrix}
        \cos(\theta_i)\cos(\varphi_i)\\[0.5ex]
        \cos(\theta_i)\sin(\varphi_i)\\[0.5ex]
        -\sin(\theta_i)
    \end{pmatrix},
    \qquad
    \hat{\textbf{e}}_i^2 = 
    \begin{pmatrix}
        -\sin(\varphi_i)\\[0.5ex]
        \cos(\varphi_i)\\[0.5ex]
        0
    \end{pmatrix},
    \qquad
    \hat{\textbf{e}}_i^3 = \hat{\textbf{n}}_i.
\end{equation}
The local axes follow the classical spin direction and therefore vary from site to site. To express the spin Hamiltonian in terms of the local spin components, we relate them to the global components through a rotation matrix. The columns of $R_i$ are the local vectors written in global coordinates,
\begin{equation}
\label{sm:eq:local rotation matrix}
    R_i=\begin{pmatrix}
        \hat{\textbf{e}}_i^1 & \hat{\textbf{e}}_i^2 & \hat{\textbf{n}}_i
    \end{pmatrix}.
\end{equation}
We denote the spin components in the local frame by $\Tilde{\textbf{S}}_i$, such that the global spins become
\begin{equation}
\label{sm:eq:local spin rotation}
    \textbf{S}_i=R_i\Tilde{\textbf{S}}_i=\Tilde{S}_i^1 \, \hat{\textbf{e}}_i^1 + \Tilde{S}_i^2 \,  \hat{\textbf{e}}_i^2 + \Tilde{S}_i^3 \, \hat{\textbf{n}}_i.
\end{equation}
The Holstein--Primakoff transformation lets us write the leading quantum fluctuations around the classical spin direction as
\begin{equation}
\label{sm:eq:HP transformation}
    \Tilde{S}_i^3 = S-a_i^\dag a_i,
    \quad
    \Tilde{S}_i^+ = \sqrt{2S} a_i,
    \quad
    \Tilde{S}_i^- = \sqrt{2S} a_i^\dag,
\end{equation}
where $\Tilde{S}_i^\pm=\Tilde{S}_i^1\pm i\Tilde{S}_i^2$, and $a_i^{\left(\dag\right)}$ is the annihilation (creation) operator for a local Holstein--Primakoff boson. 
The exchange and DMI terms in \eqref{sm:eq:Spin Hamiltonian} both couple spins on nearest-neighbor bonds. To express these in terms of the local spin frames at sites \textit{i} and \textit{j}, we first write their combined contribution to a bond $(i,j)$ in matrix form in the global frame,
\begin{equation}
    H^s_{ij} = \textbf{S}^T_i M_{ij}\textbf{S}_j,
    \quad
    M_{ij} = -J_{ij}I_3+\begin{pmatrix}
        0 & D^z_{ij} & -D^y_{ij}
        \\
        -D^z_{ij} & 0 & D^x_{ij}
        \\
        D^y_{ij} & -D^x_{ij} & 0
    \end{pmatrix}.
\end{equation}
Substituting $\textbf{S}_i=R_i\Tilde{\textbf{S}}_i$ and $\textbf{S}_j=R_j\Tilde{\textbf{S}}_j$ into the bond Hamiltonian gives its form in the local spin frames,
\begin{equation}
    H^\mr{s}_{ij} = \Tilde{\textbf{S}}^T_i g_{ij}\Tilde{\textbf{S}}_j,
    \qquad
    g_{ij} = R_i^T M_{ij}R_j.
\end{equation}
At site \textit{i}, the longitudinal component $\Tilde{S}_i^3$ is the spin component along the classical direction $\hat{\textbf{n}}_i$. The transverse components $\Tilde{S}_i^1$ and $\Tilde{S}_i^2$ are along the two local axes perpendicular to $\hat{\textbf{n}}_i$. The matrix element $g^{33}_{ij}$ couples the longitudinal components at site \textit{i} and site \textit{j}, while the transverse components couple through the block
\begin{equation}
    g_{ij,\perp} = \begin{pmatrix}
        g^{11}_{ij} & g^{12}_{ij}
        \\[1ex]
        g^{21}_{ij} & g^{22}_{ij}
    \end{pmatrix},
\end{equation}
The remaining matrix elements $g_{ij}^{a3}$ and $g_{ij}^{3a}$ with $a=1,\,2$, couple a transverse component at one site to a longitudinal component at the other. In the Holstein--Primakoff expansion above, the transverse component is linear in the boson operators and the longitudinal component has a constant and a quadratic part. Their product therefore gives linear and cubic terms. The linear terms cancel in the full spin Hamiltonian by the stationary condition in Section~\ref{sm:sec:Gnd}, and we drop the cubic terms in the quadratic approximation.
The longitudinal contribution to the bond is
\begin{equation}
    g^{33}_{ij}\,\Tilde{S}_i^3\,\Tilde{S}_j^3 = g^{33}_{ij}\left(S-a_i^\dag a_i\right)\left(S-a_j^\dag a_j\right).
\end{equation}
The constant term contributes to the classical energy, and the product of the two boson number operators is quartic. The quadratic contribution is given by
\begin{equation}
    \left(g^{33}_{ij}\,\Tilde{S}_i^3\,\Tilde{S}_j^3\right)^{(2)}= -Sg^{33}_{ij}\left(a_i^\dag a_i+a_j^\dag a_j\right).
\end{equation}
To obtain the transverse contribution, we use $\Tilde{S}_i^\pm = \Tilde{S}_i^1\pm i \Tilde{S}_i^2$ and collect the two local transverse components into a vector,
\begin{equation}
    \Tilde{\textbf{S}}_{i,\perp}=\begin{pmatrix}
        \Tilde{S}^1_i \\[1ex]
        \Tilde{S}^2_i
    \end{pmatrix}
    =
    \sqrt{\frac S2}\left(\boldsymbol{\eta}a_i+\boldsymbol{\eta}^*a_i^\dag\right),
    \qquad
    \boldsymbol{\eta}  = \begin{pmatrix}
        1
        \\
        -i
    \end{pmatrix}.
\end{equation}
Both components are linear in the boson operators at this order. Substituting them at the two ends of the bond gives terms that transfer a boson between the sites and terms that create or annihilate a pair of bosons,
\begin{equation}
    H_{ij,\perp} = \Tilde{\textbf{S}}^T_{i,\perp}g_{ij,\perp}\Tilde{\textbf{S}}_{j,\perp}=S\left(t_{ij}a_i^\dag a_j+t_{ij}^*a_j^\dag a_i+p_{ij}a_i^\dag a_j^\dag + p_
    {ij}^*a_ia_j\right),
\end{equation}
where the hopping and pair-creation amplitudes are
\begin{align}
    t_{ij}&=\frac12\boldsymbol{\eta}^\dag g_{ij,\perp}\boldsymbol{\eta},
    \\
    p_{ij}&=\frac12\boldsymbol{\eta}^\dag g_{ij,\perp}\boldsymbol{\eta}^*.
\end{align}
Adding the longitudinal and transverse contributions gives the quadratic Hamiltonian for the bond $(i,j)$
\begin{equation}
\label{sm:eq:Quadratic bond}
    H_{ij}^{s,(2)}=S\left[-g^{33}_{ij}\left(a_i^\dag a_i+a_j^\dag a_j\right) + t_{ij}a_i^\dag a_j + t_{ij}^*a_j^\dag a_i + p_{ij}a_i^\dag a_j^\dag + p_{ij}^*a_ia_j\right].
\end{equation}
The full quadratic spin Hamiltonian contains these terms summed over all nearest-neighbor bonds, together with the quadratic contribution from the on-site anisotropy.
The classical spin texture repeats from one magnetic unit cell to the next. We label each site by $(\textbf{R},\mu)$, where $\textbf{R}$ gives the position of the magnetic unit cell and $\mu$ labels the site within it. We Fourier transform the cell coordinate $\textbf{R}$ while keeping the site index $\mu$,
\begin{equation}
    a_{\textbf{R},\mu} = \frac{1}{\sqrt{N_\mr{c}}}\sum_\textbf{q} e^{-i\textbf{q}\cdot\textbf{R}}a_{\textbf{q},\mu},
\end{equation}
where $N_\mr{c}$ is the number of magnetic unit cells.
Fourier transforming the quadratic bond terms and the on-site anisotropy gives the momentum-space Hamiltonian
\begin{equation}
    H_\mr{s}^{(2)} = S\sum_\textbf{q}\sum_{\mu\nu}\left[A_{\mu\nu}(\textbf{q})a_{\textbf{q}\mu}^\dag a_{\textbf{q}\nu}+\frac12 B_{\mu\nu}(\textbf{q})a_{\textbf{q}\mu}^\dag a_{-\textbf{q},\nu}^\dag + \frac12 B_{\mu\nu}^*(\textbf{q})a_{-\textbf{q},\nu} a_{\textbf{q}\mu}\right].
\end{equation}
The factor of $1/2$ accounts for the two equivalent orderings of each boson pair in the sum. To construct \textit{A} and \textit{B}, we label the bonds by \textit{b}, counting each bond once per magnetic unit cell. For bond \textit{b} connecting $i=(\textbf{R}, \mu_b)$ to $j=(\textbf{R}+\boldsymbol{\delta}_b, \nu_b)$, we write $g_b^{33} = g_{ij}^{33}$, $t_b=t_{ij}$, and $p_b=p_{ij}$. Including the on-site anisotropy gives
\begin{align}
\notag
    A_{\mu\nu}(\textbf{q})&=\sum_b\left[-g^{33}_b\left(\delta_{\mu,\mu_b}\delta_{\nu,\mu_b}+\delta_{\mu,\nu_b}\delta_{\nu,\nu_b}\right)+t_b e^{-i\textbf{q}\cdot\boldsymbol{\delta}_b}\delta_{\mu,\mu_b}\delta_{\nu,\nu_b}+t_b^*e^{i\textbf{q}\cdot\boldsymbol{\delta}_b}\delta_{\mu,\nu_b}\delta_{\nu,\mu_b}\right]
    \\
    &\quad+K_z\left[2\cos^2(\theta_\mu)-\sin^2(\theta_\mu)\right]\delta_{\mu,\nu},
    \\
    B_{\mu\nu}(\textbf{q})&=\sum_b p_b\left[e^{-i\textbf{q}\cdot\boldsymbol{\delta}_b}\delta_{\mu,\mu_b}\delta_{\nu,\nu_b}+e^{i\textbf{q}\cdot\boldsymbol{\delta}_b}\delta_{\mu,\nu_b}\delta_{\nu,\mu_b}\right]-K_z\sin^2(\theta_\mu)\delta_{\mu,\nu}.
\end{align}
The terms proportional to $K_z$ follow from expressing the global \textit{z} component in the local frame, $S^z_i=\cos(\theta_i)\Tilde{S}_i^3-\sin(\theta_i)\Tilde{S}_i^1$.
To write the number-conserving and pair terms in a single matrix, we group the annihilation operators at $\textbf{q}$ and the creation operators at $-\textbf{q}$ into the Nambu spinor
\begin{equation}
    \boldsymbol{\Psi}_\textbf{q} = \begin{pmatrix}
        a_{\textbf{q},1}&\ldots& a_{\textbf{q},N_s}&a_{-\textbf{q},1}^\dag&\ldots&a_{-\textbf{q},N_s}^\dag
    \end{pmatrix}^T,
\end{equation}
where $N_s$ is the number of sites in the magnetic unit cell. With this ordering, the quadratic spin Hamiltonian takes the Bogoliubov-de Gennes form, up to an additive constant,
\begin{align}
\label{sm:eq:magnon BdG}
    H_\mr{s} = \frac12\sum_\textbf{q}\boldsymbol{\Psi}_\textbf{q}^\dag H_m(\textbf{q})\boldsymbol{\Psi}_\textbf{q},
    \qquad
    H_m(\textbf{q})=S\begin{pmatrix}
        A(\textbf{q}) & B(\textbf{q})
        \\[0.5ex]
        B^\dag(\textbf{q}) & A^T(-\textbf{q})
    \end{pmatrix}.
\end{align}
The diagonal blocks give the number-conserving terms at $\textbf{q}$ and $-\textbf{q}$, with the transpose in the lower block accounting for the reversed order of the creation and annihilation, respectively. The factor $1/2$ compensates for the doubling of operators in the Nambu representation.
The transformation of magnon operators must preserve the bosonic commutation relations. The annihilation and creation entries of $\boldsymbol{\Psi}_\textbf{q}$ give opposite signs in the commutator matrix,
\begin{equation}
    \left[\boldsymbol{\Psi}_\textbf{q},\boldsymbol{\Psi}_{\textbf{q}'}^\dag\right]=\delta_{\textbf{q}\textbf{q}'}\Sigma_z,
    \qquad
    \Sigma_z=\begin{pmatrix}
        I_{N_s}&0
        \\
        0& -I_{N_s}
    \end{pmatrix}.
\end{equation}
These signs enter the Heisenberg equation of motion, $i\partial_t\boldsymbol{\Psi}_\textbf{q}=\Sigma_zH_\mr{m}(\textbf{q})\boldsymbol{\Psi}_\textbf{q}$, in units with $\hbar=1$. Substituting a solution with time dependence $e^{-i\omega_{\textbf{q}\lambda}t}$ into the equation of motion gives this eigenvalue problem
\begin{equation}
\label{sm:eq:magnon eigenvalue problem}
    \Sigma_z H_m(\textbf{q})\textbf{t}_{\textbf{q}\lambda} = \omega_{\textbf{q}\lambda}\textbf{t}_{\textbf{q}\lambda},
\end{equation}
where $\textbf{t}_{\textbf{q}\lambda}$ is a $2N_s$-component eigenvector. Its upper and lower halves give the mode amplitudes in the annihilation and creation parts of $\boldsymbol{\Psi}_\textbf{q}$, respectively. The $N_s$ positive eigenvalues $\omega_{\textbf{q}\lambda}$ are the magnon energies at $\textbf{q}$. The remaining eigenvalues are $-\omega_{-\textbf{q},\lambda}$, and correspond to magnon creation operators at $-\textbf{q}$. We collect the eigenvectors as columns of $T(\textbf{q})$, placing those with positive eigenvalues first and those with negative eigenvalues last. To preserve the bosonic commutation relations, we choose the columns to satisfy
\begin{equation}
    T^\dag(\textbf{q})\Sigma_zT(\textbf{q}) = \Sigma_z.
\end{equation}
With this ordering, $T(\textbf{q})$ expresses the original Holstein--Primakoff operators in terms of the magnon operators through
\begin{equation}
\label{sm:eq:magnon transformation}
    \boldsymbol{\Psi}_\textbf{q} = T(\textbf{q})\boldsymbol{\Gamma}_\textbf{q},\qquad \boldsymbol{\Gamma}_\textbf{q} = \begin{pmatrix}
        \alpha_{\textbf{q}1}&\dots&\alpha_{\textbf{q}N_s}&\alpha_{-\textbf{q},1}^\dag&\dots&\alpha_{-\textbf{q},N_s}^\dag
    \end{pmatrix}^T.
\end{equation}
Here $\alpha_{\textbf{q},\lambda}$ annihilates a magnon in branch $\lambda$ at momentum $\textbf{q}$. Substituting this transformation into the quadratic spin Hamiltonian gives,
\begin{equation}
\label{sm:eq:diagonal magnon Hamiltonian}
    H_\mr{s}^{(2)} = \sum_{\textbf{q}\lambda}\omega_{\textbf{q}\lambda}\alpha_{\textbf{q}\lambda}^\dag\alpha_{\textbf{q}\lambda}.
\end{equation}

\section{Normal-state electrons}
\label{sm:sec:Elec}
The normal-state electron Hamiltonian contains terms describing nearest-neighbor hopping, next-nearest-neighbor hopping, the chemical potential, and the static exchange field produced by the classical spin texture, respectively,
\begin{equation}
\label{sm:eq:normal electron real space}
    H_\mr{el}=-t\sum_{\langle i,j\rangle\sigma}c_{i\sigma}^\dag c_{j\sigma}-t_2\sum_{\langle\langle i,j\rangle\rangle\sigma}c_{i\sigma}^\dag c_{j\sigma}
    -\mu\sum_{i\sigma}c_{i\sigma}^\dag c_{i\sigma}-J_\mr{sd}S\sum_{i\sigma\sigma'}\hat{\textbf{n}}_i\cdot c_{i\sigma}^\dag \boldsymbol{\sigma}_{\sigma\sigma'}c_{i\sigma'}.
\end{equation}
Here $c_{i\sigma}^{(\dag)}$ annihilates (creates) an electron at site \textit{i} with spin $\sigma$.
The sums over $\langle i,j\rangle$ and $\langle\langle i,j\rangle\rangle$ count each nearest- and next-nearest-neighbor bond once, respectively. The vector $\boldsymbol{\sigma}$ contains the Pauli matrices in the global spin basis.
The static exchange field is periodic under translations of the magnetic unit cell. Using the site labels $(\textbf{R},\mu)$, where $\textbf{R}$ points to the unit cell and $\mu$ labels the sites within it, we Fourier transform the cell coordinate $\textbf{R}$ while keeping the site and spin indices,
\begin{equation}
\label{sm:eq:electron Fourier transformation}
    c_{\textbf{R}\mu\sigma}=\frac{1}{\sqrt{N_\mr{c}}}\sum_\textbf{k}e^{-i\textbf{k}\cdot\textbf{R}}c_{\textbf{k}\mu\sigma},
\end{equation}
where $N_\mr{c}$ is the number of unit cells in the system.
The electron Hamiltonian can then be written as
\begin{equation}
\label{sm:eq:electron Bloch Hamiltonian}
    H_\mr{el}=\sum_\textbf{k}\textbf{c}_\textbf{k}^\dag h_\mr{e}(\textbf{k})\textbf{c}_\textbf{k},
\end{equation}
where $\textbf{c}_\textbf{k}$ collects the two spins components at each of the $N_s$ sites in the magnetic unit cell,
\begin{equation}
    \textbf{c}_\textbf{k}=\begin{pmatrix}
        c_{\textbf{k}1\uparrow}&c_{\textbf{k}1\downarrow}&\cdots&c_{\textbf{k}N_s\uparrow}&c_{\textbf{k}N_s\downarrow}
    \end{pmatrix}^T.
\end{equation}
Let $t_{\mu\nu}(\boldsymbol{\delta})$ denote the hopping amplitude between site $\mu$ in cell $\textbf{R}$ and site $\nu$ in cell $\textbf{R}+\boldsymbol{\delta}$. It then equals $t$ on nearest-neighbor bonds, $t_2$ on next-nearest-neighbor bonds, and zero otherwise. The displacement $\boldsymbol{\delta}$ determines the Fourier phase, giving the matrix elements
\begin{equation}
\label{sm:eq:electron Bloch matrix}
    \left[h_\mr{e}(\textbf{k})\right]_{\mu\sigma,\nu\sigma'}=-\sum_{\boldsymbol{\delta}}t_{\mu\nu}(\boldsymbol{\delta})e^{-i\textbf{k}\cdot\boldsymbol{\delta}}\delta_{\sigma\sigma'}-\mu\delta_{\mu\nu}\delta_{\sigma\sigma'}-J_\mr{sd}S\delta_{\mu\nu}\left[\hat{\textbf{n}}_\mu\cdot\boldsymbol{\sigma}\right]_{\sigma\sigma'}.
\end{equation}
Here $\hat{\textbf{n}}_\mu$ is the classical spin direction at site $\mu$.
At each momentum $\textbf{k}$, we diagonalize the matrix $h_\mr{e}(\textbf{k})$ using a unitary matrix $U(\textbf{k})$,
\begin{equation}
\label{sm:eq:electron diagonalization}
    h_\mr{e}(\textbf{k})U(\textbf{k})=U(\textbf{k})\xi(\textbf{k}).
\end{equation}
The diagonal entries of $\xi(\textbf{k})$ are the energies $\xi_{\textbf{k}n}$ of the $2N_s$ electron bands. The $n^\mr{th}$ column of $U(\textbf{k})$ gives the amplitudes $U_{\mu\sigma,n}(\textbf{k})$ of the corresponding eigenstate on each site $\mu$ and spin $\sigma$. Projecting onto these eigenvectors gives the band operators,
\begin{equation}
\label{sm:eq:electron band transformation}
    d_{\textbf{k}n}=\sum_{\mu\sigma}U_{\mu\sigma,n}^*(\textbf{k})c_{\textbf{k}\mu\sigma},\qquad
    c_{\textbf{k}\mu\sigma}=\sum_nU_{\mu\sigma,n}(\textbf{k})d_{\textbf{k}n}.
\end{equation}
In this basis, the normal-state Hamiltonian is diagonal,
\begin{equation}
\label{sm:eq:diagonal electron Hamiltonian}
    H_\mr{el}=\sum_{\textbf{k}n}\xi_{\textbf{k}n}d_{\textbf{k}n}^\dag d_{\textbf{k}n}.
\end{equation}
Since $h_\mr{e}(\textbf{k})$ already includes the chemical-potential term, the Fermi surface is given by $\xi_{\textbf{k}n}=0$.

\section{Electron-Magnon interaction}
\label{sm:sec:EL-mag}
The interaction between the localized spins and the itinerant electrons stems from the on-site exchange term,
\begin{equation}
\label{sm:eq:H sd}
    H_\mr{sd} = -J_\mr{sd}\sum_{\textbf{R}\mu}\textbf{S}_{\textbf{R}\mu}\cdot \textbf{c}_{\textbf{R}\mu}^\dag\boldsymbol{\sigma}\textbf{c}_{\textbf{R}\mu},
    \qquad
    \textbf{c}_{\textbf{R}\mu} = \begin{pmatrix}
        c_{\textbf{R}\mu\uparrow}
        \\
        c_{\textbf{R}\mu\downarrow}
    \end{pmatrix}.
\end{equation}
The classical part of $\textbf{S}_{\textbf{R}\mu}$ gives the static exchange field already included in Section~\ref{sm:sec:Elec}. To obtain the coupling to spin fluctuations, we use the local spin basis and the Holstein--Primakoff expansion from Section~\ref{sm:sec:HP}. The global spin operator then takes the form
\begin{equation}
    \textbf{S}_{\textbf{R}\mu}=S\hat{\textbf{n}}_\mu+\sqrt{\frac{S}{2}}\left[a_{\textbf{R}\mu}\left(\hat{\textbf{e}}_\mu^1-i\hat{\textbf{e}}_\mu^2\right)+a_{\textbf{R}\mu}^\dag\left(\hat{\textbf{e}}_\mu^1+i\hat{\textbf{e}}_\mu^2\right)\right]-a_{\textbf{R}\mu}^\dag a_{\textbf{R}\mu}\hat{\textbf{n}}_\mu.
\end{equation}
Substituting the spin expansion into the on-site exchange gives
\begin{equation}
    H_\mr{sd} = H_\mr{sd}^{(0)}+H_\mr{em}^{(1)}+H_\mr{em}^{(2)},
\end{equation}
where the superscript counts the number of Holstein--Primakoff boson operators in each term. The term $H_\mr{sd}^{(0)}$ is the static exchange already included in the normal-state Hamiltonian of Section~\ref{sm:sec:Elec}. The transverse spin components give the term linear in the boson operators. Defining the local spin matrices
\begin{equation}
    M_\mu^{\pm}=\left(\hat{\textbf{e}}_\mu^1\pm i\hat{\textbf{e}}_\mu^2\right)\cdot\boldsymbol{\sigma},
\end{equation}
we obtain 
\begin{equation}
    H_\mr{em}^{(1)} = -J_\mr{sd}\sqrt{\frac S2}\sum_{\textbf{R}\mu}\left[a_{\textbf{R}\mu} c_{\textbf{R}\mu}^\dag M_\mu^-c_{\textbf{R}\mu}
    + a_{\textbf{R}\mu}^\dag c_{\textbf{R}\mu}^\dag M_\mu^+c_{\textbf{R}\mu}\right].
\end{equation}
The longitudinal fluctuations change the exchange field by a term proportional to the local boson number,
\begin{equation}
    H_\mr{em}^{(2)} = J_\mr{sd} \sum_{\textbf{R}\mu}a_{\textbf{R}\mu}^\dag a_{\textbf{R}\mu}c_{\textbf{R}\mu}^\dag \left(\hat{\textbf{n}}_\mu\cdot\boldsymbol{\sigma}\right) c_{\textbf{R}\mu}.
\end{equation}
To leading order in the spin fluctuations, we only consider the linear term $H_\mr{em}^{(1)}$. The Bogoliubov transformation is linear in the boson operators, so this term describes electrons scattering by either the emission or absorption of a single magnon. We use the Fourier transformations from Sections~\ref{sm:sec:HP} and \ref{sm:sec:Elec} to write the linear coupling in momentum space,
\begin{equation}
\label{sm:eq:electron magnon momentum}
    H_\mr{em}^{(1)}=-J_\mr{sd}\sqrt{\frac{S}{2N_\mr{c}}}\sum_{\textbf{k}\textbf{q}\mu}\left[
    a_{\textbf{q}\mu}c_{\textbf{k}+\textbf{q},\mu}^\dag M_\mu^-c_{\textbf{k}\mu}
    +a_{-\textbf{q},\mu}^\dag c_{\textbf{k}+\textbf{q},\mu}^\dag M_\mu^+c_{\textbf{k}\mu}
    \right].
\end{equation}
To express the coupling in the electron band basis, we use the eigenvectors $U(\textbf{k})$ from Section~\ref{sm:sec:Elec}. The exchange acts on site $\mu$, so the matrix element involves the spin amplitudes of the incoming state $(\textbf{k},n')$ and the outgoing state $(\textbf{k}+\textbf{q},n)$ at that same site. We write this matrix element as
\begin{equation}
\label{sm:eq:local spin band matrix}
    A_{\mu,nn'}^\pm(\textbf{k}+\textbf{q},\textbf{k})
    =\sum_{\sigma\sigma'}U_{\mu\sigma,n}^*(\textbf{k}+\textbf{q})
    \left(M_\mu^\pm\right)_{\sigma\sigma'}
    U_{\mu\sigma',n'}(\textbf{k}).
\end{equation}
To express the boson operators in terms of magnons, we use $\boldsymbol{\Psi}_\textbf{q}=T(\textbf{q})\Gamma_\textbf{q}$ from Section~\ref{sm:sec:HP}. The operators $a_{\textbf{q}\mu}$ and $a_{-\textbf{q},\mu}^\dag$ occupy entries $\mu$ and $N_s+\mu$ of $\boldsymbol{\Psi}_\textbf{q}$, so the corresponding rows of $T(\textbf{q})$ give
\begin{align}
\label{sm:eq:a in terms magnon}
    a_{\textbf{q}\mu}
    &=\sum_{\lambda=1}^{N_s}\left[
    T_{\mu,\lambda}(\textbf{q})\alpha_{\textbf{q}\lambda}
    +T_{\mu,N_s+\lambda}(\textbf{q})\alpha_{-\textbf{q},\lambda}^\dag
    \right],
    \\
\label{sm:eq:a dag in terms magnon}
    a_{-\textbf{q},\mu}^\dag
    &=\sum_{\lambda=1}^{N_s}\left[
    T_{N_s+\mu,\lambda}(\textbf{q})\alpha_{\textbf{q}\lambda}
    +T_{N_s+\mu,N_s+\lambda}(\textbf{q})\alpha_{-\textbf{q},\lambda}^\dag
    \right].
\end{align}
Using $A^\pm_{\mu,nn'}$ for the electron matrix elements and collecting the terms that create or annihilate a magnon gives
\begin{equation}
\label{sm:eq:electron magnon band coupling}
    H_\mr{em}^{(1)}
    =\sum_{\textbf{k}\textbf{q}}\sum_{\lambda nn'}
    \left[
    g_{\lambda nn'}^{(1)}(\textbf{k}+\textbf{q},\textbf{k})
    \alpha_{-\textbf{q},\lambda}^\dag
    +
    g_{\lambda nn'}^{(2)}(\textbf{k}+\textbf{q},\textbf{k})
    \alpha_{\textbf{q}\lambda}
    \right]
    d_{\textbf{k}+\textbf{q},n}^\dag d_{\textbf{k}n'}.
\end{equation}
The vertices $g^{(1)}$ and $g^{(2)}$ describe magnon emission at $-\textbf{q}$ and absorption at $\textbf{q}$, respectively. Both processes scatter the electron from $(\textbf{k},n')$ to $(\textbf{k}+\textbf{q},n)$.
Summing the contributions from all sites $\mu$ in the magnetic unit cell gives the vertices
\begin{align}
\label{sm:eq:electron magnon vertices}
    g_{\lambda nn'}^{(1)}(\textbf{k}+\textbf{q},\textbf{k})
    &=-J_\mr{sd}\sqrt{\frac{S}{2N_\mr{c}}}
    \sum_\mu\left[
    T_{\mu,N_s+\lambda}(\textbf{q})
    A_{\mu,nn'}^-(\textbf{k}+\textbf{q},\textbf{k})
    +T_{N_s+\mu,N_s+\lambda}(\textbf{q})
    A_{\mu,nn'}^+(\textbf{k}+\textbf{q},\textbf{k})
    \right],
    \\
    g_{\lambda nn'}^{(2)}(\textbf{k}+\textbf{q},\textbf{k})
    &=-J_\mr{sd}\sqrt{\frac{S}{2N_\mr{c}}}
    \sum_\mu\left[
    T_{\mu,\lambda}(\textbf{q})
    A_{\mu,nn'}^-(\textbf{k}+\textbf{q},\textbf{k})
    +T_{N_s+\mu,\lambda}(\textbf{q})
    A_{\mu,nn'}^+(\textbf{k}+\textbf{q},\textbf{k})
    \right].
\end{align}

\subsection{Effective electron--electron interaction}
Two electrons can interact by exchanging a virtual magnon. To derive this interaction, we apply a Schrieffer--Wolff transformation~\cite{SW66} to $H=H_0+\eta H_1$, where $H_0=H_\mr{el}+H_\mr{s}^{(2)}$ describes the uncoupled electrons and magnons, $H_1=H_\mr{em}^{(1)}$ is the electron-magnon coupling, and $\eta$ counts powers of this coupling. The transformed Hamiltonian is
\begin{equation}
    H'=e^{-\eta\mathcal{S}}He^{\eta\mathcal{S}},
\end{equation}
where $\mathcal{S}$ is an anti-Hermitian generator. Expanding to second order in $\eta$ gives
\begin{equation}
\label{sm:eq:SW expansion}
\begin{aligned}
    H'=&H_0+\eta\left(H_1+\big[H_0,\mathcal{S}\big]\right)
    +\eta^2\left(\big[H_1,\mathcal{S}\big]
    +\frac12\big[[H_0,\mathcal{S}],\mathcal{S}\big]\right)
    +\mathcal{O}\left(\eta^3\right).
\end{aligned}
\end{equation}
We choose $\mathcal{S}$ so that the coupling linear in $\eta$ cancels,
\begin{equation}
\label{sm:eq:SW condition}
    H_1+[H_0,\mathcal{S}]=0.
\end{equation}
The commutator with $H_0$ multiplies each operator product in $H_1$ by the energy change of the corresponding process. We therefore write $\mathcal{S}$ in terms of the same operator products, with coefficients $x_{\lambda nn'}$ and $y_{\lambda nn'}$ for magnon emission and absorption, respectively,
\begin{equation}
\label{sm:eq:SW ansatz}
\begin{aligned}
    \mathcal{S}
    =\sum_{\textbf{k}\textbf{q}}\sum_{\lambda nn'}
    \left[
    x_{\lambda nn'}(\textbf{k},\textbf{q})
    g_{\lambda nn'}^{(1)}(\textbf{k}+\textbf{q},\textbf{k})
    \alpha_{-\textbf{q},\lambda}^\dag
    +
    y_{\lambda nn'}(\textbf{k},\textbf{q})
    g_{\lambda nn'}^{(2)}(\textbf{k}+\textbf{q},\textbf{k})
    \alpha_{\textbf{q}\lambda}
    \right]
    d_{\textbf{k}+\textbf{q},n}^\dag d_{\textbf{k}n'}.
\end{aligned}
\end{equation}
The energy change includes the change in electron energy and the energy of the emitted or absorbed magnon, namely,
\begin{equation}
\label{sm:eq:SW commutators}
\begin{aligned}
    \left[
    H_0,
    \alpha_{-\textbf{q},\lambda}^\dag
    d_{\textbf{k}+\textbf{q},n}^\dag d_{\textbf{k}n'}
    \right]
    &=
    \left(
    \xi_{\textbf{k}+\textbf{q},n}
    -\xi_{\textbf{k}n'}
    +\omega_{-\textbf{q},\lambda}
    \right)
    \alpha_{-\textbf{q},\lambda}^\dag
    d_{\textbf{k}+\textbf{q},n}^\dag d_{\textbf{k}n'},
    \\
    \left[
    H_0,
    \alpha_{\textbf{q}\lambda}
    d_{\textbf{k}+\textbf{q},n}^\dag d_{\textbf{k}n'}
    \right]
    &=
    \left(
    \xi_{\textbf{k}+\textbf{q},n}
    -\xi_{\textbf{k}n'}
    -\omega_{\textbf{q}\lambda}
    \right)
    \alpha_{\textbf{q}\lambda}
    d_{\textbf{k}+\textbf{q},n}^\dag d_{\textbf{k}n'}.
\end{aligned}
\end{equation}
Matching the coefficients of these operator products in the cancellation condition and solving for $x$ and $y$ gives
\begin{equation}
\begin{aligned}
    1+x_{\lambda nn'}(\textbf{k},\textbf{q})
    \left(
    \xi_{\textbf{k}+\textbf{q},n}
    -\xi_{\textbf{k}n'}
    +\omega_{-\textbf{q},\lambda}
    \right)&=0,
    \\
    1+y_{\lambda nn'}(\textbf{k},\textbf{q})
    \left(
    \xi_{\textbf{k}+\textbf{q},n}
    -\xi_{\textbf{k}n'}
    -\omega_{\textbf{q}\lambda}
    \right)&=0.
\end{aligned}
\end{equation}
Solving for the coefficients yields
\begin{equation}
\label{sm:eq:SW coefficients}
\begin{aligned}
    x_{\lambda nn'}(\textbf{k},\textbf{q})
    &=
    \frac{1}{
    \xi_{\textbf{k}n'}
    -\xi_{\textbf{k}+\textbf{q},n}
    -\omega_{-\textbf{q},\lambda}},
    \\[0.5ex]
    y_{\lambda nn'}(\textbf{k},\textbf{q})
    &=
    \frac{1}{
    \xi_{\textbf{k}n'}
    +\omega_{\textbf{q}\lambda}
    -\xi_{\textbf{k}+\textbf{q},n}}.
\end{aligned}
\end{equation}
With this choice for $\mathcal{S}$, the transformed Hamiltonian becomes
\begin{equation}
\label{sm:eq:SW effective Hamiltonian}
    H'= H_0+\frac12\eta^2[H_1,\mathcal{S}]
    +\mathcal{O}\left(\eta^3\right).
\end{equation}
The terms with four electron operators in the commutator describe the interaction mediated by virtual magnon exchange. We retain the process that scatter a zero-total-momentum pair from $(\textbf{k}',n')$, $(-\textbf{k}',m')$ to $(\textbf{k},n)$, $(-\textbf{k},m)$. Setting $\eta=1$, the resulting pairing term has the form
\begin{equation}
\label{sm:eq:effective elec-elec}
    H_{\mr{pair}}=\sum_{\textbf{k}\textbf{k}'}\sum_{nm,n'm'}V_{\textbf{k}\textbf{k}'}^{nm,n'm'}d_{\textbf{k}n}^\dag d_{-\textbf{k},m}^\dag d_{-\textbf{k}',m'}d_{\textbf{k}',n'}.
\end{equation}
The first electron gains momentum $\textbf{q}=\textbf{k}-\textbf{k}'$, while the second gains $-\textbf{q}$. We evaluate the interaction with all four electron states on the Fermi surface, where the electronic energy differences in the denominator vanish. Using $\omega_{-\textbf{q},\lambda} = \omega_{\textbf{q},\lambda}$, we obtain
\begin{equation}
\label{sm:eq:SW interaction}
    V_{\textbf{k}\textbf{k}'}^{nm,n'm'}=-\frac12\sum_\lambda\frac{1}{\omega_{\textbf{k}-\textbf{k}',\lambda}}\Big[
    g_{\lambda nn'}^{(1)}(\textbf{k},\textbf{k}')g_{\lambda mm'}^{(2)}(-\textbf{k},-\textbf{k}')
    +g_{\lambda nn'}^{(2)}(\textbf{k},\textbf{k}')g_{\lambda mm'}^{(1)}(-\textbf{k},-\textbf{k}')\Big].
\end{equation}
Exchanging the electrons in either the incoming or the outgoing pair reverses the sign of the four-electron operator. Combining the four orderings gives the antisymmetrized interaction
\begin{equation}
\label{sm:eq:Pauli antisymmetrized interaction}
\begin{aligned}
    \shortbar{V}_{\textbf{k}\textbf{k}'}^{nm,n'm'}=\frac12\Big(&
    V_{\textbf{k}\textbf{k}'}^{nm,n'm'}
    +V_{-\textbf{k},-\textbf{k}'}^{mn,m'n'}
    -V_{-\textbf{k},\textbf{k}'}^{mn,n'm'}
    -V_{\textbf{k},-\textbf{k}'}^{nm,m'n'}\Big).
\end{aligned}
\end{equation}
We can see that this interaction obeys the antisymmetric relations
\begin{equation}
    \shortbar{V}_{\textbf{k}\textbf{k}'}^{nm,n'm'}
    =\shortbar{V}_{-\textbf{k},-\textbf{k}'}^{mn,m'n'}
    =-\shortbar{V}_{-\textbf{k},\textbf{k}'}^{mn,n'm'}
    =-\shortbar{V}_{\textbf{k},-\textbf{k}'}^{nm,m'n'}.
\end{equation}
With all of this, the resulting effective electron Hamiltonian is given by
\begin{equation}
\label{sm:eq:effective electron Hamiltonian}
    H_\mr{eff}=\sum_{\textbf{k}n}\xi_{\textbf{k}n}d_{\textbf{k}n}^\dag d_{\textbf{k}n}+\frac12\sum_{\textbf{k}\textbf{k}'}\sum_{nm,n'm'}\shortbar{V}_{\textbf{k}\textbf{k}'}^{nm,n'm'}d_{\textbf{k}n}^\dag d_{-\textbf{k},m}^\dag d_{-\textbf{k}',m'}d_{\textbf{k}',n'}.
\end{equation}

\section{Superconductivity}
\label{sm:sec:SC}
To derive the multiband gap equation, we apply a mean-field approximation to the pairing term of the effective electron Hamiltonian. For the pair of states $(\textbf{k},n)$ and $(-\textbf{k},m)$, we define the mean pair amplitude
\begin{equation}
    b_{\textbf{k}nm}= \Big\langle d_{-\textbf{k},m}d_{\textbf{k}n}\Big\rangle,
    \qquad 
    b_{\textbf{k}nm}^*= \left\langle d_{\textbf{k}n}^\dag d_{-\textbf{k},m}^\dag\right\rangle.
\end{equation}
We write the pair operator as
$d_{-\textbf{k},m}d_{\textbf{k}n}=b_{\textbf{k}nm}+\delta b_{\textbf{k}nm}$, where $\delta b_{\textbf{k}nm}=d_{-\textbf{k},m}d_{\textbf{k}n}-\langle d_{-\textbf{k},m}d_{\textbf{k}n}\rangle$. Keeping terms up to first order in the fluctuations gives
\begin{equation}
    d_{\textbf{k}n}^\dag d_{-\textbf{k},m}^\dag d_{-\textbf{k}',m'}d_{\textbf{k}',n'} \approx
    b_{\textbf{k}'n'm'} d_{\textbf{k}n}^\dag d_{-\textbf{k},m}^\dag 
    + b_{\textbf{k}nm}^*  d_{-\textbf{k}',m'}d_{\textbf{k}',n'} -  b_{\textbf{k}'n'm'}b_{\textbf{k}nm}^*.
\end{equation}
We define the superconducting gap by
\begin{equation}
    \Delta_{\textbf{k}nm} = -\sum_{\textbf{k}'n'm'}\shortbar{V}_{\textbf{k}\textbf{k}'}^{nm,n'm'}b_{\textbf{k}'n'm'}.
\end{equation}
Substituting the decoupled pairing term into $H_\mr{eff}$ gives
\begin{equation}
    H_\mr{MF} = \sum_{\textbf{k}n}\xi_{\textbf{k}n}d_{\textbf{k}n}^\dag d_{\textbf{k}n} + \frac12\sum_{\textbf{k}nm}\Delta_{\textbf{k}nm}b_{\textbf{k}nm}^* -\frac12\sum_{\textbf{k}nm}\left(\Delta_{\textbf{k}nm}d_{\textbf{k}n}^\dag d_{-\textbf{k},m}^\dag + \Delta_{\textbf{k}nm}^* d_{-\textbf{k},m} d_{\textbf{k},n}\right),
\end{equation}
Exchanging the two states in a pair changes the sign of the pair operator. The antisymmetry of $\shortbar{V}$ gives the same property for the gap,
\begin{equation}
    b_{-\textbf{k}mn} = -b_{\textbf{k}nm}, \qquad \Delta_{-\textbf{k}mn}=-\Delta_{\textbf{k}nm}.
\end{equation}
The pairing shell extends from $-\omega_\mr{M}$ to $\omega_\mr{M}$ relative to the Fermi level, where $\omega_\mr{M}$ is the maximum magnon energy. We include a pair $(\textbf{k},n)$, $(-\textbf{k},m)$ only if both states lie within this shell,
\begin{equation}
    |\xi_{\textbf{k}n}|\leq\omega_\mr{M}, \qquad  |\xi_{-\textbf{k},m}|\leq\omega_\mr{M}
\end{equation}
Several normal-state bands can contribute to the pairing at the same momentum. Let $n_1,\ldots,n_\textbf{k}$ label the bands within the pairing shell at $\textbf{k}$, and let $m_1,\ldots,m_{-\textbf{k}}$ label those at $-\textbf{k}$. The Nambu spinor collects the annihilation operators at $\textbf{k}$ and the creation operators at $-\textbf{k}$,
\begin{equation}
    \boldsymbol{\Psi}_\textbf{k} = \begin{pmatrix}
        d_{\textbf{k}n_1} & \dots & d_{\textbf{k}N_\textbf{k}} & d_{-\textbf{k},m_1}^\dag & \dots d_{-\textbf{k},m_{N_{-\textbf{k}}}}^\dag
    \end{pmatrix}^T.
\end{equation}
In this basis, the normal-state matrix is diagonal. The gap matrix connects the states at $\textbf{k}$ to those at $-\textbf{k}$, so it has dimensions $N_\textbf{k}\times N_{-\textbf{k}}$. The matrix elements are
\begin{align}
    \left[\xi(\textbf{k})\right]_{rr'} = \xi_{\textbf{k}n_r}\delta_{rr'},
    \qquad
    \left[\Delta(\textbf{k})\right]_{rs} = \Delta_{\textbf{k}n_rm_s}.
\end{align}
With the sign convention for the gap in $H_\mr{MF}$, the Bogoliubov-de Gennes matrix is
\begin{equation}
\label{sm:eq:BdG matrix}
    \mathcal{H}_\mr{BdG}(\textbf{k}) = \begin{pmatrix}
        \xi(\textbf{k}) & -\Delta(\textbf{k})
        \\[0.5ex]
        -\Delta^\dag(\textbf{k}) & -\xi^T(-\textbf{k})
    \end{pmatrix}.
\end{equation}
Within the pairing shell, the mean-field Hamiltonian takes the form
\begin{align}
    H_\mr{MF} = C(\Delta, \Delta^*) + \frac12\sum_\textbf{k}\boldsymbol{\Psi}_\textbf{k}^\dag \mathcal{H}_\mr{BdG}(\textbf{k})\boldsymbol{\Psi}_\textbf{k}.
\end{align}
The factor of $1/2$ compensates for the Nambu doubling. The term $C(\Delta)$ includes the constant from normal ordering the hole block and the constant from the mean-field decoupling. Using $\Delta=-\shortbar{V}b$, we obtain
\begin{align}
    C(\Delta) = \frac12\sum_\textbf{k}\mr{Tr}\left[\xi(\textbf{k})\right]-\frac12\sum_{\textbf{k}\textbf{k}'}\sum_{nm,n'm'}\Delta^*_{\textbf{k}nm}\left(\shortbar{V}^{-1}\right)_{\textbf{k}\textbf{k}'}^{nm,n'm'}\Delta_{\textbf{k}'n'm'}.
\end{align}
We diagonalize the BdG Hamiltonian as
\begin{equation}
    \mathcal{H}_\mr{BdG}(\textbf{k})W_\textbf{k}=W_\textbf{k}E_\textbf{k},
\end{equation}
where the columns of $W_\textbf{k}$ are the orthonormal eigenvectors and $E_\textbf{k}$ is diagonal with the $N_\textbf{k} + N_{-\textbf{k}}$ BdG eigenvalues. The corresponding mean-field grand potential is
\begin{equation}
\label{sm:eq:Grand potential def}
    \Omega(\Delta,\Delta^*) = C(\Delta, \Delta^*) -\frac{1}{2\beta}\sum_\textbf{k}\mr{Tr}\left(\ln\left[1+e^{-\beta\mathcal{H}_\mr{BdG}(\textbf{k})}\right]\right),
\end{equation}
where $\beta=1/k_BT$. 
The gap equation follows from requiring the grand potential to be stationary. We treat $\Delta$ and $\Delta^*$ as independent variables when taking partial derivatives. Since $\Omega$ is real, the two stationary conditions are complex conjugates, so we only need the derivative with respect to $\Delta^*$. 

For a scalar function \textit{g}, the diagonalization of the BdG Hamiltonian gives
\begin{equation}
\label{sm:eq:Function of H}
    g\left(\mathcal{H}_\mr{BdG}(\textbf{k})\right)=W_\textbf{k}g\left(E_\textbf{k}\right)W_\textbf{k}^\dag,
\end{equation}
where $g(E_\textbf{k})$ is diagonal with entries $g(E_{\textbf{k}\nu})$. To differentiate the trace term in the grand potential, we use the identity
\begin{equation}
    \pdv{}{x}\mr{Tr}\, g\left[\mathcal{H}(x)\right] = \mr{Tr}\left[g'\left[\mathcal{H}(x)\right]\pdv{\mathcal{H}}{x}\right].
\end{equation}
Here $g(E)=\mr{ln}(1+e^{-\beta E})$, so $g'(E)=-\beta f(E)$, where $f(E)=(e^{\beta E}+1)^{-1}$ is the Fermi--Dirac distribution.
Differentiating $C(\Delta,\Delta^*)$ with respect to the conjugate gap gives
\begin{equation}
\label{sm:eq:C derivative}
    \pdv{C}{\Delta_{\textbf{k}n_rm_s}^*} = -\frac12\sum_{\textbf{k}'n'm'}\left(\shortbar{V}^{-1}\right)_{\textbf{k}\textbf{k}'}^{n_rm_s,n'm'}\Delta_{\textbf{k}'n'm'}.
\end{equation}
In the trace term, $\Delta^*_{\textbf{k}n_rm_s}$ appears with a minus sign in the lower-left block of $H_\mr{BdG}(\textbf{k})$, at row $N_\textbf{k}+s$ and column \textit{r}. Using $g'(E)=-\beta f(E)$, we obtain
\begin{equation}
\label{sm:eq:Trace}
    \pdv{}{\Delta_{\textbf{k}n_rm_s}^*}\left[-\frac{1}{2\beta}\sum_{\textbf{k}'}\mr{Tr}\,\ln\left(1+e^{-\beta\mathcal{H}_\mr{BdG}(\textbf{k}')}\right)\right] = \frac12\mr{Tr}\,\left[f\left(\mathcal{H}_\mr{BdG}(\textbf{k})\right)\pdv{\mathcal{H}_\mr{BdG}(\textbf{k})}{\Delta^*_{\textbf{k}n_rm_s}}\right].
\end{equation}
The matrix derivative on the right-hand side has elements
\begin{equation}
    \left[\pdv{\mathcal{H}_\mr{BdG}(\textbf{k})}{\Delta_{\textbf{k}n_rm_s}^*}\right]_{ij}=-\delta_{i,N_\textbf{k}+s}\delta_{j,r}.
\end{equation}
Using the general trace identity
\begin{equation}
    \mr{Tr}(AB)=\sum_{ij}A_{ij}B_{ji}
\end{equation}
for two general matrices \textit{A}, and \textit{B}, we find
\begin{equation}
    \mr{Tr}\left[f\left(\mathcal{H}_\mr{BdG}(\textbf{k})\right)\pdv{\mathcal{H}_\mr{BdG}(\textbf{k})}{\Delta^*_{\textbf{k}n_rm_s}}\right] = -\left[f\left(\mathcal{H}_\mr{BdG}(\textbf{k})\right)\right]_{r,N_\textbf{k}+s}.
\end{equation}
Substituting this result into \eqref{sm:eq:Trace} and adding the derivative of $C$ from \eqref{sm:eq:C derivative} gives
\begin{equation}
\label{sm:eq:Derivative grand potential}
    \pdv{\Omega}{\Delta_{\textbf{k}n_rm_s}^*} = -\frac12\left[\,\sum_{\textbf{k}'n'm'}\left(\shortbar{V}^{-1}\right)_{\textbf{k}\textbf{k}'}^{n_rm_s,n'm'}\Delta_{\textbf{k}'n'm'}
    + \left[f\left(\mathcal{H}_\mr{BdG}(\textbf{k})\right)\right]_{r,N_\textbf{k}+s}\right].
\end{equation}
In the basis that diagonalizes the BdG Hamiltonian, the Nambu spinor is
\begin{equation}
    \boldsymbol{\Gamma}_\textbf{k} = W_\textbf{k}^\dag\boldsymbol{\Psi}_\textbf{k},
\end{equation}
and the thermal expectation values are
\begin{equation}
    \left\langle\Gamma^\dag_{\textbf{k}\mu}\Gamma_{\textbf{k}\nu}\right\rangle =\delta_{\mu\nu}f(E_{\textbf{k}\nu}).
\end{equation}
Transforming back with $\boldsymbol{\Psi}_\textbf{k}=W_\textbf{k}\boldsymbol{\Gamma}_\textbf{k}$ gives
\begin{equation}
    \left\langle\Psi_{\textbf{k},j}^\dag\Psi_{\textbf{k},i}\right\rangle
    =
    \sum_{\nu=1}^{N_\textbf{k}+N_{-\textbf{k}}}(W_\textbf{k})_{i\nu}(W_\textbf{k})^*_{j\nu}f(E_{\textbf{k}\nu})
    =
    \left[f(\mathcal{H}_\mr{BdG}(\textbf{k}))\right]_{ij}.
\end{equation}
For $i=r$ and $j=N_\textbf{k}+s$, the Nambu entries are $\Psi_{\textbf{k}r}=d_{\textbf{k}n_r}$ and $\Psi_{\textbf{k},N_\textbf{k}+s}^\dag=d_{-\textbf{k},m_s}$. The matrix element in the derivative of the grand potential is therefore the pair amplitude,
\begin{equation}
\label{sm:eq:f as W}
    \left[f\left(\mathcal{H}_\mr{BdG}(\textbf{k})\right)\right]_{r,N_\textbf{k}+s} = \left\langle d_{-\textbf{k}m_s} d_{\textbf{k}n_r}\right\rangle 
    =b_{\textbf{k}n_rm_s}.
\end{equation}
Setting the derivative, \eqref{sm:eq:Derivative grand potential}, to zero
and inserting \eqref{sm:eq:f as W}, we solve for the gap as
\begin{equation}
\label{sm:eq:General gap equation sum}
    \Delta_{\textbf{k}n_rm_s}=-\sum_{\textbf{k}'r's'}\shortbar{V}_{\textbf{k}\textbf{k}'}^{n_rm_s,n_{r'}m_{s'}}\sum_{\nu=1}^{N_{\textbf{k}'}+N_{-\textbf{k}'}}W_{\textbf{k}', r', \nu}W^*_{\textbf{k}',N_{\textbf{k}'}+s',\nu}f(E_{\textbf{k}'\nu}).
\end{equation}
where $n'_r$ and $m_s'$ label the bands at $\textbf{k}'$ and $-\textbf{k}'$, respectively. The BdG eigenvalues and eigenvectors on the right-hand side depend on the gap, so this equation must be solved self-consistently.
\begin{equation}
    \frac{1}{N_\mr{c}}\sum_{\textbf{k}'}\rightarrow\frac{1}{A_\mr{BZ}}\int_\mr{BZ}\dd^2k',
\end{equation}
where $A_\mr{BZ}$ is the area of the Brillouin zone. For each incoming band pair $n'$, $m'$, let $S_{n'm'}$ denote the momenta allowed by the pairing-shell restriction. The gap equation becomes
\begin{equation}
    \Delta_{\textbf{k}nm} = -\frac{N_\mr{c}}{A_\mr{BZ}}\sum_{n'm'}\int_{S_{n'm'}}\dd^2k' \,\shortbar{V}_{\textbf{k}\textbf{k}'}^{nm,n'm'}\left\langle d_{-\textbf{k}',m'}d_{\textbf{k}'n'}\right\rangle.
\end{equation}
For each incoming band $n'$, we use the band energy $\xi=\xi_{\textbf{k}'n'}$ and the arc length \textit{s} along the constant-energy contour
\begin{equation}
    C_{n'}(\xi)=\{\textbf{k}'\in\mr{BZ}|\xi_{\textbf{k}'n'}=\xi\},
\end{equation}
as integration variables. The energy gradient is perpendicular to the contour, so a displacement $\dd k'_\perp$ along the gradient changes the energy by $\dd\xi=|\nabla_{\textbf{k}'}\xi_{\textbf{k}'n'}|\dd k'_\perp$. The momentum-space element is therefore
\begin{equation}
    \dd^2k=\frac{\dd s\,\dd\xi}{\left|\nabla_\textbf{k}\xi_{\textbf{k}n}\right|}.
\end{equation}
For a general integrand $F(\textbf{k}')$, the integral over allowed momenta is
\begin{equation}
    \int_{\mathcal{S}_{nm}}\!\!\dd^2k'F(\textbf{k}') = \int_{-\omega_\mr{M}}^{\omega_\mr{M}}\!\!\dd\xi\oint_{C_{n'}(\xi)}\frac{\dd s}{\left|\nabla_{\textbf{k}'}\xi_{\textbf{k'}n'}\right|}F(\textbf{k}')\Theta(\omega_\mr{M}-|\xi_{-\textbf{k}',m'}|).
\end{equation}
The limits of the energy integral impose the cutoff for $(\textbf{k}',n')$, and the step function $\Theta$ imposes it for $(-\textbf{k}',m')$.

For the projection onto the Fermi surface, we include the parts of the shell whose contours can be followed continuously to $\xi=0$. Let $\textbf{k}_\mr{F}$ and $\textbf{k}'_\mr{F}$ be the points associated with $\textbf{k}$ and $\textbf{k}'$ on the Fermi surface of bands $n$ and $n'$, respectively. Since the shell is narrow compared with the electronic bandwidth, we approximate the gap and the interaction as constant along the direction normal to the Fermi surface,
\begin{equation}
    \Delta_{\textbf{k}nm}\approx \Delta_{\textbf{k}_\mr{F}nm},\qquad \shortbar{V}^{nm,n'm'}_{\textbf{k}\textbf{k}'}\approx \shortbar{V}^{nm,n'm'}_{\textbf{k}_\mr{F}\textbf{k}_\mr{F}'}.
\end{equation}
We keep the full electronic dispersions when evaluating the pair amplitude at $\textbf{k}'=\textbf{k}'(\xi,s)$. The projected gap equation is
\begin{equation}
\label{sm:eq:Gap FS}
    \Delta_{\textbf{k}_\mr{F}nm} = -\frac{N_\mr{c}}{A_\mr{BZ}}\sum_{n'm'}\int_{-\omega_\mr{M}}^{\omega_\mr{M}}\!\!\dd\xi\oint_{C_{n'}(\xi)}\frac{\dd s}{\left|\nabla_{\textbf{k}'}\xi_{\textbf{k}'n'}\right|}\shortbar{V}^{nm,n'm'}_{\textbf{k}_\mr{F}\textbf{k}_\mr{F}'}\left\langle d_{-\textbf{k}',m'}d_{\textbf{k}'n'}\right\rangle\Theta\left(\omega_\mr{M}-|\xi_{-\textbf{k}',m'}|\right).
\end{equation}
We will now consider two different temperature limits. First is the critical temperature, at this temperature, the gap amplitude is small, so we can linearize the gap equation. The second limit is the zero-temperature limit, here we need to solve the full nonlinear equation self-consistently.

\subsection{Linearization close to the critical temperature}
As the temperature approaches $T_c$ from below, the gap tends to zero. In the normal state, the BdG matrix is diagonal in the band basis, so the pair amplitude vanishes. To find its leading contribution, we separate the BdG matrix into its normal-state and pairing parts,
\begin{gather}
    \mathcal{H}_\mr{BdG}(\textbf{k}) = \mathcal{H}_\mr{BdG}^{(0)}(\textbf{k}) +\mathcal{H}_\Delta(\textbf{k}),
    \\[1ex]
    \mathcal{H}_\mr{BdG}^{(0)}(\textbf{k})=\begin{pmatrix}
        \xi(\textbf{k}) & 0
        \\
        0 & -\xi^T(-\textbf{k})
    \end{pmatrix},
    \qquad
    \mathcal{H}_\Delta(\textbf{k})=\begin{pmatrix}
        0 & -\Delta(\textbf{k})
        \\
        -\Delta^\dag(\textbf{k}) & 0
    \end{pmatrix}.
\end{gather}
Because $\mathcal{H}_\mr{BdG}^{(0)}(\textbf{k})$ is diagonal, the term linear in the gap in a given electron-hole entry of $f(\mathcal{H}_\mr{BdG})$ depends only on the gap element connecting those two states. We can therefore calculate that term for the pair $(\textbf{k},n)$, $(-\textbf{k},m)$ using the matrix
\begin{equation}
    h_{\textbf{k}nm} = \begin{pmatrix}
        \xi_{\textbf{k}n} & -\Delta_{\textbf{k}_\mr{F}nm}
        \\[0.5ex]
        -\Delta_{\textbf{k}_\mr{F}nm}^* & -\xi_{-\textbf{k},m}
    \end{pmatrix}.
\end{equation}
Its two eigenvalues are
\begin{equation}
    E^\pm_{\textbf{k}nm} = \frac{\xi_{\textbf{k}n}-\xi_{-\textbf{k},m}}{2}\pm\sqrt{\left(\frac{\xi_{\textbf{k}n}+\xi_{-\textbf{k},m}}{2}\right)^2+|\Delta_{\textbf{k}_\mr{F}nm}|^2}.
\end{equation}
For this two-state matrix, we choose a linear function $p(E)=\alpha+\gamma E$ that equals $f(E)$ at both eigenvalues. Since $p(E)$ and $f(E)$ agree at both eigenvalues of $h_{\textbf{k}nm}$, they act identically on both eigenvectors. This gives
\begin{equation}
    \gamma_{\textbf{k}nm} = \frac{f(E^+_{\textbf{k}nm})-f(E^-_{\textbf{k}nm})}{E^+_{\textbf{k}nm}-E^-_{\textbf{k}nm}}.
\end{equation}
Only the term $\gamma h_{\textbf{k}nm}$ contributes to the off-diagonal entry, giving
\begin{equation}
    \left[f(h_{\textbf{k}nm})\right]_{12} = \gamma_{\textbf{k}nm}\left[h_{\textbf{k}nm}\right]_{12} = \Delta_{\textbf{k}_\mr{F}nm}\frac{f(E_{\textbf{k}nm}^-)-f(E_{\textbf{k}nm}^+)}{E_{\textbf{k}nm}^+-E_{\textbf{k}nm}^-}.
\end{equation}
To first order in the gap, we evaluate the ratio at the normal-state energies $\xi_{\textbf{k}n}$ and $-\xi_{\textbf{k},m}$. Using $f(-E)=1-f(E)$, the pair amplitude becomes
\begin{equation}
    \left\langle d_{-\textbf{k},m}d_{\textbf{k}n}\right\rangle =\chi_{\textbf{k}nm}(T)\Delta_{\textbf{k}_\mr{F}nm}+\mathcal{O}\left(|\Delta|^2\right),\qquad \chi_{\textbf{k}nm}(T) = \frac{1-f(\xi_{\textbf{k}n})-f(\xi_{-\textbf{k},m})}{\xi_{\textbf{k}n}+\xi_{-\textbf{k},m}}.
\end{equation}
Here $\chi_{\textbf{k}nm}(T)$ is the normal-state pair susceptibility. When the denominator vanishes, the ratio has the continuous limit $-f'(\xi_{\textbf{k}n})$. 

Substituting the linearized pair amplitude into the projected gap equation \eqref{sm:eq:Gap FS} gives
\begin{equation}
\label{sm:eq:Linear BCS}
    \Delta_{\textbf{k}_\mr{F}nm} = -\frac{N_\mr{c}}{A_\mr{BZ}}\sum_{n'm'}\int_{-\omega_\mr{M}}^{\omega_\mr{M}}\!\!\dd\xi\oint_{C_{n'}(\xi)}\frac{\dd s}{\left|\nabla_{\textbf{k}'}\xi_{\textbf{k}'n'}\right|}\shortbar{V}_{\textbf{k}_\mr{F}\textbf{k}_\mr{F}'}^{nm,n'm'}\chi_{\textbf{k}'n'm'}(T)\Delta_{\textbf{k}'_\mr{F}n'm'}\Theta\left(\omega_\mr{M}-\left|\xi_{-\textbf{k}',m'}\right|\right).
\end{equation}
This equation is linear and homogeneous in the gap and can be written as
\begin{equation}
\label{sm:eq:Pairing Kernel}
    \Delta = L(T)\Delta.
\end{equation}
In the numerical calculation, $i=(\textbf{k}_{\mr{F},i}, n_i, m_i)$ labels a Fermi-surface momentum and the two band indices of the pair. We collect the integration weights and the pair susceptibility for each incoming pair \textit{j} into $S_j(T)$, giving
\begin{equation}
    \Delta_i = -\sum_j\shortbar{V}_{ij}S_j(T)\Delta_j.
\end{equation}
We define $D_{ij}(T)=\sqrt{S_i(T)}\delta_{ij}$ and $\varphi=D(T)\Delta$. The gap equation then becomes
\begin{equation}
\label{sm:eq:Hermitian matrix equation}
    \varphi=K(T)\varphi,\qquad K(T)=-D(T)\shortbar{V}D(T).
\end{equation}
Since $\shortbar{V}$ is Hermitian, $K(T)$ is Hermitian and has the same eigenvalues as $L(T)$. The critical temperature is determined by lowering the temperature until the largest eigenvalue reaches one,
\begin{equation}
    \lambda_\mr{max}(T) = 1.
\end{equation}
The corresponding eigenvector gives the momentum- and band-dependence of the gap at the transition through $\Delta=D^{-1}\varphi$, up to an overall normalization.

\subsection{Zero temperature limit}
At zero temperature, the Fermi--Dirac distribution reduces to a step function,
\begin{equation}
    \lim_{T\rightarrow0^+}f(E) = \Theta(-E),
\end{equation}
with $\Theta(0)=1/2$. The pair amplitude is then
\begin{equation}
    \left\langle d_{-\textbf{k},m} d_{\textbf{k}n}\right\rangle_{T=0} = \sum_{\nu=1}^{N_{\textbf{k}}+N_{-\textbf{k}}}W_{\textbf{k}n\nu} W^*_{\textbf{k},N_\textbf{k}+m,\nu}\Theta(-E_{\textbf{k}\nu}).
\end{equation}
Substituting this expression into the projected gap equation \eqref{sm:eq:Gap FS}, gives
\begin{equation}
\label{sm:eq:Zero temp BCS}
\begin{aligned}
    \Delta_{\textbf{k}_\mr{F}nm} = &-\frac{N_\mr{c}}{A_\mr{BZ}}\sum_{n'm'}\int_{-\omega_\mr{M}}^{\omega_\mr{M}}\!\!\dd\xi\oint_{C_{n'}(\xi)}\frac{\dd s}{\left|\nabla_{\textbf{k}'}\xi_{\textbf{k}'n'}\right|}\shortbar{V}_{\textbf{k}_\mr{F}\textbf{k}_\mr{F}'}^{nm,n'm'}\sum_{\nu=1}^{N_{\textbf{k}'}+N_{-\textbf{k}'}}W_{\textbf{k}'n'\nu} W^*_{\textbf{k}',N_{\textbf{k}'}+m',\nu}
    \\[1ex]
    &\times\Theta(-E_{\textbf{k}'\nu})\Theta(\omega_\mr{M}-|\xi_{-\textbf{k}',m'}|).
\end{aligned}
\end{equation}
Here $r'$ and $s'$ are the positions of bands $n'$ and $m'$ in the electron and hole blocks of the Nambu spinor at $\textbf{k}'$, respectively. The eigenvalues and eigenvectors of the full multiband BdG matrix depend on the gap. So we solve this equation self-consistently. We use the momentum- and band-dependent gap found at $T_c$ as the starting point of the iteration. Section~\ref{sm:sec:Numerical gap} describes the numerical procedure.

\section{Critical temperature enhancement}
\label{sm:sec:Tc}
\begin{figure}[tb]
    \centering
    \includegraphics[width=0.7\linewidth]{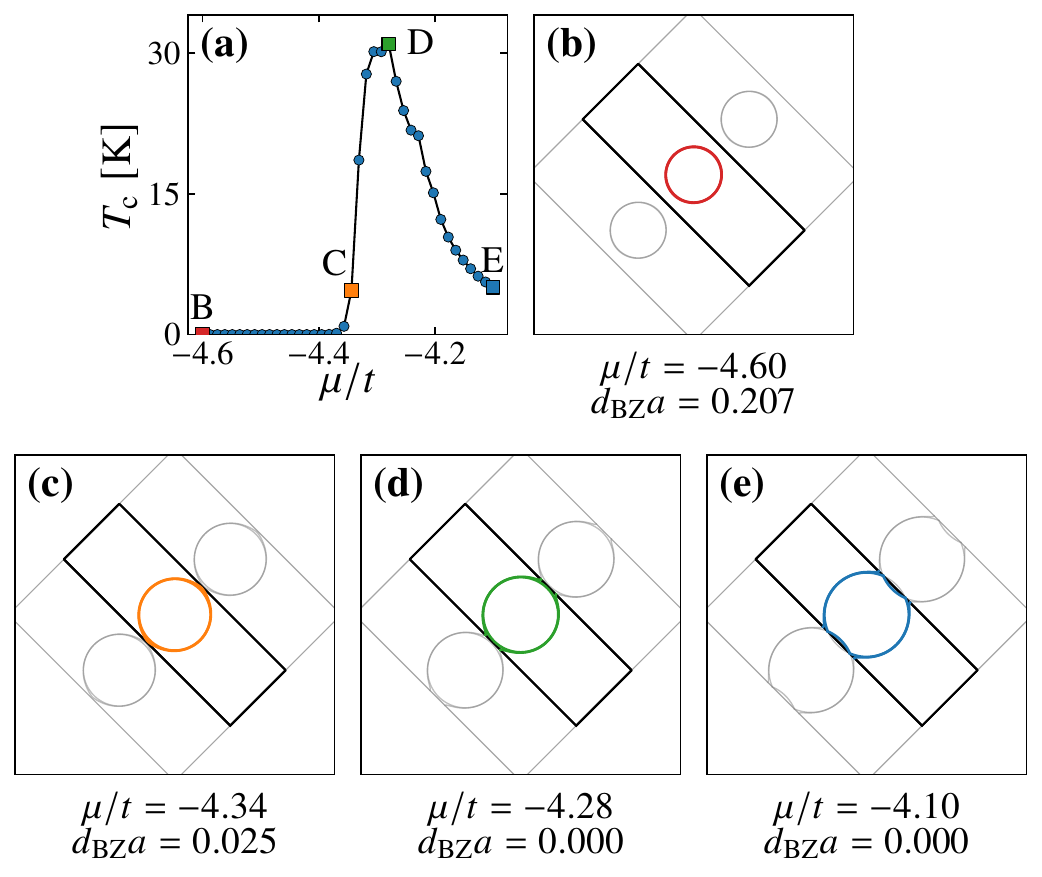}
    \caption{
    \textbf{(a)} Critical temperature $T_c$ as a function of $\mu/t$, with cases \textit{B}-\textit{E} marked.
    \textbf{(b)}-\textbf{(e)} Corresponding Fermi surfaces for cases \textit{B}-\textit{E}, respectively. The black rectangle outlines the first magnetic Brillouin zone. The gray outlines and curves show neighboring zones, and the translated copies of the Fermi surface. The quantity $d_\mr{BZ}a$ is the dimensionless minimum distance from the Fermi surface to the boundary of the first zone.
    }
    \label{sm:fig:Tc and FS}
\end{figure}

The critical temperature rises sharply and then falls as the chemical potential increases, we can see this in \figref{sm:fig:Tc and FS}{a}, where we also mark four cases along the curve. Case \textit{B} has a low $T_c$, \textit{C} lies on the rise, \textit{D} is at the maximum, and \textit{E} lies on the falling side. The Fermi surface in \textit{B} lies well inside the magnetic Brillouin zone. It approaches the boundary in \textit{C} and reaches it in \textit{D} and \textit{E}, as shown in \figref{sm:fig:Tc and FS}{b}-\figref{sm:fig:Tc and FS}{e}. Near the boundary, Umklapp scattering can connect states on opposite sides of the zone through small reduced momentum transfers. To investigate the role of these transfers, we evaluate the pairing kernel at the same temperature $T_*=10$~K in all four cases and denote its leading eigenvalue by $\lambda=\lambda_\mr{max}(T_*)$. For scattering from a pair at $\textbf{k}'$ to a pair at $\textbf{k}$, the reduced direct and exchange momentum transfers are
\begin{figure}[tb]
    \centering
    \includegraphics[width=0.7\linewidth]{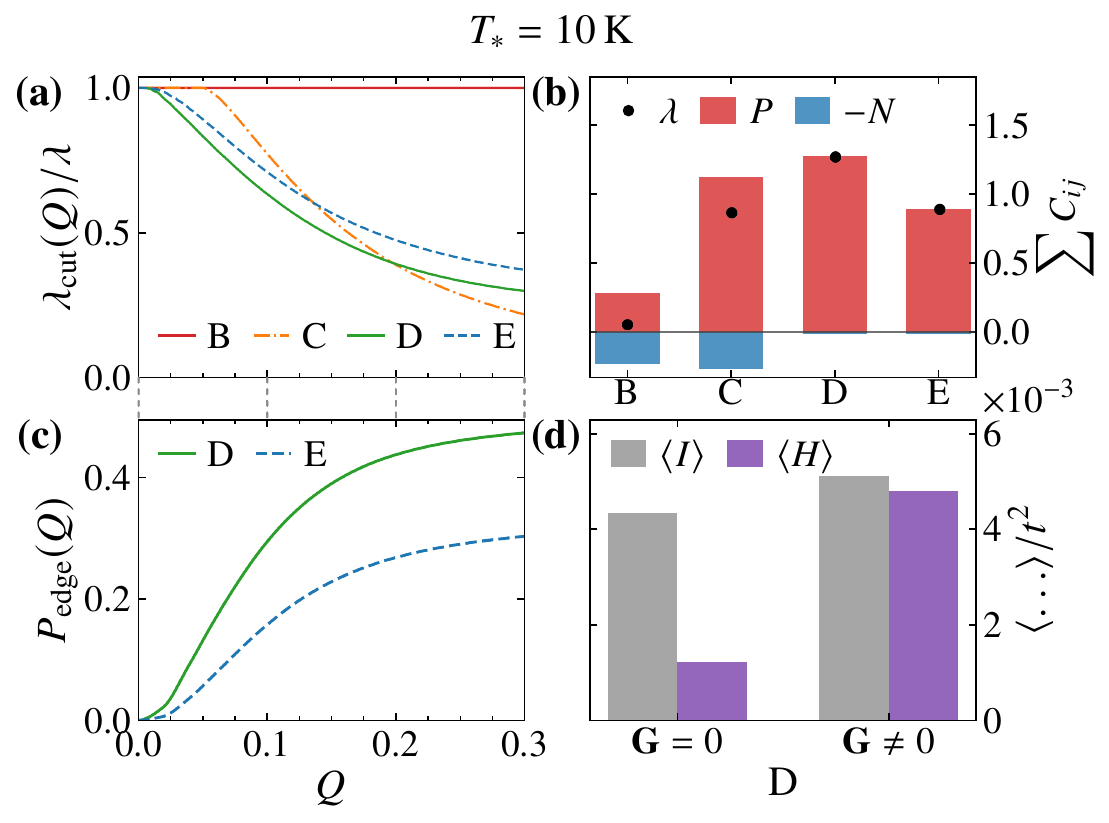}
    \caption{
    Pairing diagnostics for cases \textit{B}-\textit{E} in \figref{sm:fig:Tc and FS}{a} at a common temperature $T_*=10$~K.
    \textbf{(a)} Ratio of the leading eigenvalues of the modified and full kernels, $\lambda_\mr{cut}(Q)/\lambda$. A kernel element is set to zero if either its direct or exchange transfer crosses the zone boundary and satisfies $a|\textbf{q}|<Q$. 
    \textbf{(b)} The bars show the summed positive and negative contributions \textit{P} and $-N$. The black dots mark $\lambda=P-N$.
    \textbf{(c)} Cumulative positive contribution $P_\mr{edge}(Q)$ in \textit{D} and \textit{E} from elements whose direct transfers cross the zone boundary and satisfy $a|\textbf{q}_d|<Q$.
    \textbf{(d)} Averages $\langle I\rangle/t^2$ and $\langle H\rangle/t^2$ of the squared electron-magnon coupling in case \textit{D}. The two groups, $\textbf{G}_d=0$ and $\textbf{G}_d\neq0$, are compared at matched reduced direct transfers. The quantity \textit{I} uses the sum of squared site amplitudes, while \textit{H} sums the amplitudes before taking the squared magnitude.
    }
    \label{sm:fig:Tc explanation}
\end{figure}
\begin{equation}
    \textbf{q}_\mr{d}=\textbf{k}'-\textbf{k}-\textbf{G}_\mr{d},\qquad \textbf{q}_\mr{x}=\textbf{k}'+\textbf{k}-\textbf{G}_\mr{x},
\end{equation}
where $\textbf{G}_\mr{d}$ and $\textbf{G}_\mr{x}$ are magnetic reciprocal lattice vectors that bring the transfers into the first Brillouin zone. A nonzero reciprocal shift identifies scattering across the zone boundary. For each dimensionless cutoff $Q=q_\mr{max} a$, we set a kernel element to zero if either transfer has a nonzero reciprocal shift and satisfies $a|\textbf{q}|<Q$. We then calculate the leading eigenvalue $\lambda_\mr{cut}(Q)$ of the modified kernel. \figref{sm:fig:Tc explanation}{a} shows that at $Q=0.15$, the ratio $\lambda_\mr{cut}(Q)/\lambda$ is one in \textit{B} and roughly one half in \textit{C}-\textit{E}.

The increase in $\lambda$ can come from larger positive contributions or weaker cancellation by negative ones. To distinguish these effects, we use the normalized leading eigenvector $\varphi$ of the full kernel, with $\sum_i|\varphi_i|^2=1$. Multiplying the eigenvalue equation \eqref{sm:eq:Hermitian matrix equation},
\begin{equation}
\label{sm:eq:BCS K}
    \lambda\varphi_i=\sum_jK_{ij}\varphi_j,
\end{equation}
by $\varphi_i^*$ and summing over \textit{i} gives
\begin{equation}
    \lambda=\sum_{ij}\mr{Re}\left(\varphi_i^* K_{ij}\varphi_j\right) = \sum_{ij}C_{ij}.
\end{equation}
The sum is real because \textit{K} is Hermitian. We write the total positive contribution as \textit{P} and the magnitude of the total negative contribution as \textit{N},
\begin{equation}
    P=\sum_{ij}\mr{max}(C_{ij},0),
    \qquad
    N = \sum_{ij}\mr{max}(-C_{ij},0),
    \qquad
    \lambda = P-N.
\end{equation}
\figref{sm:fig:Tc explanation}{b} shows that the positive and negative terms nearly cancel in \textit{B}. The increase from \textit{B} to \textit{C} comes mainly from the growth of \textit{P}, and the further increase from \textit{C} to \textit{D} comes mainly from a reduction in \textit{N}. The decrease from \textit{D} to \textit{E} comes mainly from a loss of positive contributions.

To locate that loss, we group the full kernel elements by their direct transfers and sum the positive contributions whose direct transfer crosses the zone boundary,
\begin{equation}
    P_\mr{edge}(Q) =\!\!\!\!\!\!\!
    \sum_{\substack{i,j 
    \\
    \textbf{G}_{\mr{d},ij}\neq 0
    \\
    a|\textbf{q}_{\mr{d},ij}|<Q}}
    \!\!\!\!\!\!\max\left(C_{ij},0\right).
\end{equation}
From \figref{sm:fig:Tc explanation}{c}, we see that at $Q=0.30$, $P_\mr{edge}$ decreases from about $0.47$ in \textit{D} to about $0.30$ in \textit{E}, accounting for about $45\%$ of the total decrease in \textit{P}. The difference is already about $0.16$ at $Q=0.15$, so nearly all of the drop seen at $Q=0.30$ comes from the smaller reduced transfers. This decrease accompanies the change in the Fermi-surface shaper near the zone boundary in \figref{sm:fig:Tc and FS}{d} and \figref{sm:fig:Tc and FS}{e}

The results above show that Umklapp scattering with small reduced momentum transfers plays an important role in the enhancement of pairing. To understand why these processes are strong, we look at how the electron-magnon coupling builds up from the sites of the magnetic unit cell. Each site contributes an amplitude, and the relative phases determine whether these amplitudes reinforce or partly cancel one another.

The label $\nu=\pm$ distinguishes the two electrons in a Cooper pair. When the pair scatters from \textit{j} to \textit{i}, the incoming and outgoing states of the two electrons are
\begin{align*}
    (\textbf{k}_j,n_j)\rightarrow(\textbf{k}_i,n_i)\quad (\nu=+),
    \qquad
    (-\textbf{k}_j,m_j)\rightarrow(-\textbf{k}_i,m_i)\quad (\nu=-).
\end{align*}
We denote the contribution from site \textit{s} by $f^\nu_{sr\lambda}(i,j)$, where $r=1$ denotes magnon emission, $r=2$ denotes absorption, and $\lambda$ labels the magnon branch. The full vertex is the sum of these site amplitudes,
\begin{equation}
    g^\nu_{r\lambda}(i,j) = \sum_{s=1}^{N_s}f_{sr\lambda}^\nu(i,j).
\end{equation}
For each electron, we compare the sum of the squared amplitudes on individual sites $I_\nu$ with the strength obtained by summing the amplitudes before squaring, $H_\nu$,
\begin{align}
    I_\nu(i,j)&=\sum_{r\lambda}\sum_s\left|f^\nu_{sr\lambda}(i,j)\right|^2,
    \\
    H_\nu(i,j)&=\sum_{r\lambda}\left|\sum_sf^\nu_{sr\lambda}(i,j)\right|^2.
\end{align}
Their difference comes from the interference between sites as,
\begin{equation}
    \left|\sum_sf^\nu_{sr\lambda}(i,j)\right|^2 = \sum_s\left|f^\nu_{sr\lambda}(i,j)\right|^2+2\sum_{s',\,s<s'}\mr{Re}\left(f^*_s(i,j)f_{s'}(i,j)\right),
\end{equation}
gives
\begin{equation}
    H_\nu(i,j)-I_\nu(i,j)=2\sum_{r\lambda}\sum_{s',\,s<s'}\mr{Re}\left(\left[f_{sr\lambda}^\nu(i,j)\right]^*f_{s'r\lambda}(i,j)\right).
\end{equation}
A negative difference means that the site amplitudes partly cancel. To include both electrons in the comparison, we take the geometric means
\begin{equation}
    I_{ij} = \sqrt{I_+(i,j)I_-(i,j)},\qquad H_{ij} = \sqrt{H_+(i,j)H_-(i,j)}.
\end{equation}
\figref{sm:fig:Tc explanation}{d} compares $\langle I\rangle/t^2$ and $\langle H\rangle/t^2$ in case \textit{D} for processes with $\textbf{G}_d=0$ and $\textbf{G}_d\neq0$ at matched reduced direct transfers. The two groups have comparable $\langle I\rangle$. The reduction from $\langle I\rangle$ to $\langle H\rangle$ is much larger for $\textbf{G}_d=0$, indicating that the Umklapp vertices are stronger because their site amplitudes cancel less.

\section{Topological superconductivity}
\label{sm:sec:Topology}
The superconducting gap vanishes at isolated points on the Fermi surface, producing point nodes in the bulk BdG spectrum. For a given edge direction, fixing the momentum $k_\parallel$ along the edge defines a one-dimensional cut through the bulk Brillouin zone. The cut is gapless when $k_\parallel$ coincides with the projection of a bulk node. On the gapped cuts, chiral symmetry allows us to define a winding number. We therefore first identify the symmetry operations that act within a fixed-$k_\parallel$ cut.

\subsection{Symmetries of a fixed momentum cut}
In addition to $k_\parallel$, we define $k_\perp$ to be the momentum component perpendicular to the edge. A symmetry within the cut acts within the cut if it preserves $k_\parallel$, but it can still reverse $k_\perp$. 

In the Nambu basis, the particle-hole operator is $\Xi=\tau_x\mathcal{K}$, where $\tau_x$ exchanges the particle and hole blocks and \textit{K} denotes complex conjugation. It satisfies
\begin{equation}
\label{sm:eq:BdG particle hole}
    \Xi\mathcal{H}_\mr{BdG}(\textbf{k})\Xi^{-1}
    =
    -\mathcal{H}_\mr{BdG}(-\textbf{k}).
    \qquad
    \Xi^2=1,
\end{equation}
This relation follows from the particle-hole structure of the BdG matrix and the antisymmetry of the gap, $\boldsymbol\Delta(-\textbf{k})=-\boldsymbol{\Delta}^T(\textbf{k})$. A state at momentum $\textbf{k}$ and energy $E_\textbf{k}$ therefore has a particle-hole partner at $-\textbf{k}$ and energy $-E_\textbf{k}$,
\begin{equation}
    \mathcal{H}_\mr{BdG}(-\textbf{k})
    \Xi\ket{u_{\textbf{k}}}
    =
    -E_{\textbf{k}}\Xi\ket{u_{\textbf{k}}}.
\end{equation}
Since both ordinary time reversal and a translation by half a unit cell reverse the magnetic texture, the combination of these operations is a symmetry,
\begin{equation}
\label{sm:eq:magnetic time reversal}
    \Theta=\hat{t}_{\boldsymbol{\tau}}\mathcal{T},
    \qquad
    \mathcal{T}=i\sigma_y\mathcal{K}.
\end{equation}
For the superconducting state considered here,
\begin{equation}
\label{sm:eq:magnetic time reversal BdG}
    \Theta\mathcal{H}_\mr{BdG}(\textbf{k})\Theta^{-1}
    =
    \mathcal{H}_\mr{BdG}(-\textbf{k}).
\end{equation}
Both magnetic time reversal and particle-hole symmetry relate the cut at $k_\parallel$ to the cut at $-k_\parallel$. To construct operations that act within a fixed cut, we combine them with a symmetry that reverses $k_\parallel$ and leaves $k_\perp$ unchanged. For the edge parallel to $(1,1)$, the spatial part of this symmetry is a reflection about the line $x+y=c$. We choose \textit{c} so that the reflection combined with a spin rotation by $\pi$ about \textit{z} preserves the magnetic texture. We denote this combined operation by $M_{\bar{d}}$. Its spin rotation is represented by $-i\sigma_z$. The combined operation acts on the electron spinor at site $(x,y)$ as
\begin{equation}
\label{sm:eq:Manti operator}
    M_{\bar{d}}c_{(x,y)}
    M_{\bar{d}}^{-1}
    =
    -i\sigma_zc_{(c-y,c-x)}.
\end{equation}
The reflection preserves $x-y$, so it maps each boundary of the diagonal ribbon to itself.

For a gap that preserves the combined reflection, we use its BdG representation to define
\begin{equation}
    \Theta_\mr{1D}=M_{\bar{d}}\Theta,
    \qquad
    \Xi_\mr{1D}=M_{\bar{d}}\Xi.
\end{equation}
Both operations preserve $k_\parallel$ and reverse $k_\perp$. Their action on the bulk BdG Hamiltonian is
\begin{align}
    \Theta_\mr{1D}\mathcal{H}_\mr{BdG}(k_\parallel,k_\perp)\Theta_\mr{1D}^{-1}
    &=
    \mathcal{H}_\mr{BdG}(k_\parallel,-k_\perp),
    \\
    \Xi_\mr{1D}\mathcal{H}_\mr{BdG}(k_\parallel,k_\perp)\Xi_\mr{1D}^{-1}
    &=
    -\mathcal{H}_\mr{BdG}(k_\parallel,-k_\perp).
\end{align}
Their product $\Gamma=\Theta_\mr{1D}\Xi_\mr{1D}$ is unitary and leaves both momentum components fixed. It defines a chiral symmetry because it anticommutes with the BdG Hamiltonian,
\begin{equation}
\label{sm:eq:chiral symmetry topology}
    \Gamma\mathcal{H}_\mr{BdG}(k_\parallel,k_\perp)\Gamma^{-1}
    =
    -\mathcal{H}_\mr{BdG}(k_\parallel,k_\perp).
\end{equation}
At a given momentum, we choose the phase of $\Gamma$ so that $\Gamma^2=1$. Its $+1$ and $-1$ eigenspaces define the chiral basis,
\begin{equation}
    U_\Gamma\Gamma U_\Gamma^\dag
    =
    \begin{pmatrix}
        I & 0
        \\
        0 & -I
    \end{pmatrix}.
\end{equation}
The anticommutation relation means that the Hamiltonian maps each chiral eigenspace into the other. Its diagonal blocks therefore vanish in this basis,
\begin{equation}
    U_\Gamma\mathcal{H}_\mr{BdG}(k_\parallel,k_\perp)U_\Gamma^\dag
    =
    \begin{pmatrix}
        0 & q(k_\parallel,k_\perp)
        \\
        q^\dag(k_\parallel,k_\perp) & 0
    \end{pmatrix}.
\end{equation}
For a gapped cut, the two chiral eigenspaces have equal dimension and \textit{q} is a square matrix. A zero of $\mr{det}q$ would give a zero-energy state of the BdG Hamiltonian, so $\mr{det} q$ is nonzero for every $k_\perp$ on the cut.

\subsection{Momentum-resolved winding number}
On a gapped cut, the determinant of the off-diagonal block has a well-defined phase,
\begin{equation}
    \det q(k_\parallel,k_\perp)
    =
    \left|\det q(k_\parallel,k_\perp)\right|
    e^{i\varphi(k_\parallel,k_\perp)}.
\end{equation}
Along a transverse path $\mathcal{C}$ that closes in the chosen chiral basis, the determinant returns to its initial value. Its phase can change by an integer multiple of $2\pi$, which defines the winding number
\begin{equation}
\label{sm:eq:winding number}
    \nu(k_\parallel)
    =
    \frac{1}{2\pi i}
    \oint_{\mathcal{C}}
    \!\!\!\!\dd k_\perp\,
    \partial_{k_\perp}
    \ln\det q(k_\parallel,k_\perp)=
    \frac{1}{2\pi}
    \oint_{\mathcal{C}}
    \!\!\!\!\dd k_\perp\,
    \partial_{k_\perp}
    \varphi(k_\parallel,k_\perp).
\end{equation}
The coplanar magnetic texture gives an additional unitary symmetry. A spin rotation by $\pi$ about the normal $\hat{\textbf{n}}_p$ to the spin plane reverses all the local moments. Since the half-period translation $\hat{t}_{\boldsymbol{\tau}}$ also reverses the moments, the combined operation preserves the texture,
\begin{equation}
    U_{\boldsymbol{\tau}}=\hat{t}_{\boldsymbol{\tau}} R_p,\qquad
    R_p=-i\hat{\textbf{n}}_p\cdot\boldsymbol{\sigma}.
\end{equation}
For spin-$1/2$ electrons, $R_p^2=-I$, so $U_{\boldsymbol{\tau}}^2=-\hat{t}_{2\boldsymbol{\tau}}$. Its action on a normal-state Bloch eigenstate is therefore
\begin{equation}
\label{sm:eq:magnetic sector eigenvalue}
    U_{\boldsymbol{\tau}}(\textbf{k})
    \ket{u_{\textbf{k}\lambda}}
    =
    \lambda i e^{-i\textbf{k}\cdot\boldsymbol{\tau}}
    \ket{u_{\textbf{k}\lambda}},
    \qquad
    \lambda=\pm1.
\end{equation}
We call $\lambda$ the sector index. Magnetic time reversal commutes with $U_{\boldsymbol{\tau}}$ and complex conjugates its eigenvalue. It therefore maps sector $\lambda$ at $\textbf{k}$ to sector $-\lambda$ at $-\textbf{k}$.

In the Nambu basis, an electron in sector $\lambda$ at $\textbf{k}$ and a hole associated with sector $-\lambda$ at $-\textbf{k}$ carry the same $U_{\boldsymbol{\tau}}$ eigenvalue. With this pairing structure, the BdG Hamiltonian separates into two sectors at each momentum. The chiral operator commutes with the BdG representation of $U_{\boldsymbol{\tau}}$, so we can choose the chiral basis within each sector. The off-diagonal block then takes the form
\begin{equation}
    q(k_\parallel,k_\perp)
    =
    \begin{pmatrix}
        q_+(k_\parallel,k_\perp) & 0
        \\
        0 & q_-(k_\parallel,k_\perp)
    \end{pmatrix}.
\end{equation}
The sector and chiral labels must be followed continuously along the transverse momentum cut. For the edge parallel to $(1,1)$, a transverse reciprocal vector is $\textbf{G}_\perp=(\pi,-\pi)/a$. Since $e^{-i\textbf{G}_\perp\cdot\boldsymbol{\tau}}=-1$, the $U_{\boldsymbol{\tau}}$ eigenvalue satisfies
\begin{equation}
    i\lambda e^{-i(\textbf{k}+\textbf{G}_\perp)\cdot\boldsymbol{\tau}}=(-\lambda)ie^{-i\textbf{k}\cdot\boldsymbol{\tau}}.
\end{equation}
The momenta $\textbf{k}$ and $\textbf{k}+\textbf{G}_\perp$ are one transverse period apart and represent the same point in the Brillouin zone. Since $\mr{exp}(-i\textbf{G}_\perp\cdot\boldsymbol{\tau})=-1$, the sector labeled $\lambda$ at $\textbf{k}$ is labeled $-\lambda$ at $\textbf{k}+\textbf{G}_\perp$. The normalized chiral operator also changes sign, so a state with chiral eigenvalue $+1$ at one endpoint has eigenvalue $-1$ at the other. After two periods, both labels return to their starting values.

We call this two-period path at fixed $k_\parallel$ for $\mathcal{C}_\lambda$. We choose the chiral basis continuously along the path and use the same basis vectors at its two endpoints. This ensures that $\mr{det}q_\lambda$ returns to its initial value. If the BdG spectrum has no zero-energy states anywhere along the path, $\mr{det}q_\lambda$ never vanishes, and its phase defines the winding
\begin{equation}
\label{sm:eq:sector winding}
    \nu_\lambda(k_\parallel)
    =
    \frac{1}{2\pi i}
    \oint_{\mathcal{C}_\lambda}\!\!\!
    \dd k_\perp\,
    \partial_{k_\perp}
    \ln\det q_\lambda(k_\parallel,k_\perp).
\end{equation}
Since $\mr{det}q=\mr{det}q_+\mr{det}q_-$, the winding of the full block over the same two periods is $\nu_++\nu_-$.
In the weak-pairing limit, the gap can be reduced to zero away from the Fermi surface without closing the BdG gap. We choose the hole basis by applying $\Gamma$ to the electron basis. Where the gap is set to zero, $q_\lambda$ then reduces to the normal-state energy matrix, whose determinant is real and nonzero. The phase changes are therefore concentrated near the Fermi-surface crossings.

For an isolated crossing $j$ at $k_j=(k_\parallel,k_{\perp,j})$, we expand the normal-state energy along the cut,
\begin{equation}
    \xi_{\lambda,j}(k_\perp)
    \approx
    v_{\perp,j}\left(k_\perp-k_{\perp,j}\right),
\qquad
    v_{\perp,j}
    =
    \left.
    \pdv{\xi_{\lambda,j}}{k_\perp}
    \right|_j.
\end{equation}
Let $\ket{e_j}$ be the electron state at this crossing and choose the hole state as $\ket{h_j}=\Gamma(\textbf{k}_j)\ket{e_j}$. This fixes the relative phase of the two states. The anticommutation of $\Gamma$ with $\mathcal{H}_\mr{BdG}$ gives
\begin{equation}
    \bra{h_j}\mathcal{H}_\mr{BdG}\ket{e_j} = -\bra{e_j}\mathcal{H}_\mr{BdG}\ket{h_j}.
\end{equation}
Since the two matrix elements are complex conjugates, they are purely imaginary. The signed gap in this basis is the real quantity
\begin{equation}
    \Delta_{\lambda,j} = i\bra{e_j}\mathcal{H}_\mr{BdG}\ket{h_j}.
\end{equation}
An arbitrary phase on $\ket{e_j}$ also multiplies $\ket{h_j}$ by the same phase, so it cancels in this matrix element.

In the chiral basis $(\ket{e_j}\pm\ket{h_j})/\sqrt{2}$, the scalar part of $q_\lambda$ associated with the crossing is
\begin{equation}
\label{sm:eq:weak pairing local q}
    q_{\lambda,j}(k_\perp)
    \approx
    v_{\perp,j}\left(k_\perp-k_{\perp,j}\right)
    +i\Delta_{\lambda,j}.
\end{equation}
The sign of $v_{\perp,j}$ determines the direction in which this path is traversed. The sign of $\Delta_{\lambda,j}$ determines whether it passes above or below the origin. The phase change in the weak-pairing limit is
\begin{equation}
    \delta\varphi_{\lambda,j}
    =
    -\pi
    \operatorname{sgn}\left(v_{\perp,j}\right)
    \operatorname{sgn}\left(\Delta_{\lambda,j}\right).
\end{equation}
Adding these changes gives
\begin{equation}
\label{sm:eq:weak pairing winding}
    \nu_\lambda(k_\parallel)
    =
    -\frac12
    \sum_{\substack{
    j\text{ at FS crossings}\\
    \text{along }\mathcal{C}_\lambda
}}
    \operatorname{sgn}\left(v_{\perp,j}\right)
    \operatorname{sgn}\left(\Delta_{\lambda,j}\right).
\end{equation}
The sum includes the crossings encountered in both transverse periods. 
Magnetic time reversal maps a crossing in sector $\lambda$ at $\textbf{k}$ to one in sector $-\lambda$ at $-\textbf{k}$. The transverse velocity changes sign, while the signed gap is the same in the chiral convention used above. 
At fixed $k_\parallel$, shifting the path by $G_\perp$ also exchanges the sectors. This shift preserves the velocity and reverses the gap sign in the chiral basis because $\Gamma(\textbf{k}+\textbf{G}_\perp)=-\Gamma(\textbf{k})$. Thus,
\begin{equation}
    \nu_\lambda(k_\parallel) = -\nu_{-\lambda}(k_\parallel),
\end{equation}
and the sum $\nu_++\nu_-$ vanishes on this two-period path.

\subsection{Point nodes and their edge projections}
The winding can change when a transverse cut passes through a point node. At an isolated node $\textbf{k}_{\lambda i}$, the normal-state energy and the signed gap of the crossing band both vanish,
\begin{equation}
    \xi_\lambda(\textbf{k}_{\lambda i})=0,
    \qquad
    \Delta_\lambda(\textbf{k}_{\lambda i})=0.
\end{equation}
Near the node, the contribution from this band to the off-diagonal block has the form $q_{\lambda i}(\textbf{k})\approx \xi_\lambda(\textbf{k})+i\Delta_\lambda(\textbf{k})$. The topological charge is the winding of its phase around a small counterclockwise loop $\ell_{\lambda i}$ enclosing the node~\cite{Review_TSC},
\begin{figure}
    \centering
    \includegraphics[width=0.4\linewidth]{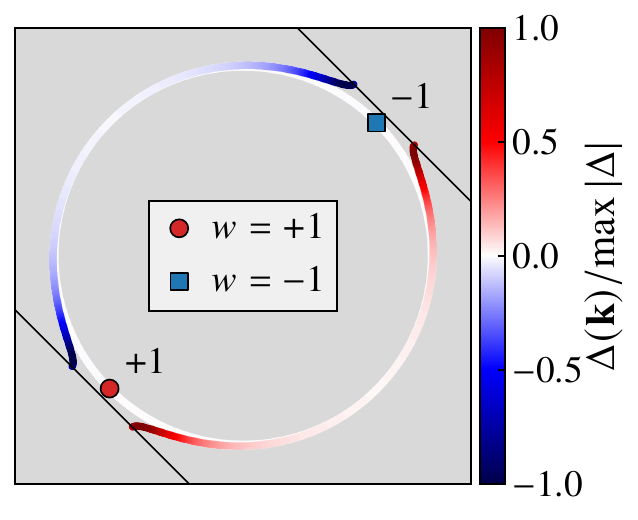}
    \caption{Topological charges of the nodes on the Fermi surface. The color shows the signed superconducting gap divided by its maximum amplitude. The red circle and blue square mark the nodes with charges $w=+1$ and $w=-1$, respectively.}
    \label{sm:fig:Topological charge}
\end{figure}
\begin{equation}
\label{sm:eq:nodal charge}
    w_{\lambda i}
    =
    \frac{1}{2\pi}
    \oint_{\ell_{\lambda i}}
    \!\!\!\dd\textbf{k}\cdot
    \nabla_\textbf{k}
    \arg\left[q_{\lambda i}(\textbf{k})\right].
\end{equation}
For a node, expanding $\xi_\lambda$ and $\Delta_\lambda$ to first order gives
\begin{equation}
    w_{\lambda i}
    =
    \operatorname{sgn}
    \left[
    \pdv{\xi_\lambda}{k_x}
    \pdv{\Delta_\lambda}{k_y}
    -
    \pdv{\xi_\lambda}{k_y}
    \pdv{\Delta_\lambda}{k_x}
    \right]_{\textbf{k}=\textbf{k}_{\lambda i}}.
\end{equation}
Fig.~\ref{sm:fig:Topological charge} shows the two nodes and their opposite topological charges. They lie near the line $k_y=k_x$ on opposite sides of the origin. For the edge parallel to $(1,1)$, the projected momenta are
\begin{equation}
    k_{\parallel,\lambda i} = \textbf{k}_{\lambda i}\cdot\hat{\textbf{e}}_\parallel, 
    \qquad
     \hat{\textbf{e}}_\parallel = \frac{\hat{\textbf{x}}+\hat{\textbf{y}}}{\sqrt{2}}.
\end{equation}
The projections are distinct, so the two nodes close the gap on different fixed-$k_\parallel$ cuts.

\section{Numerical methods}
\label{sm:sec:Numerics}
The pairing interaction uses magnon vertices at opposite momentum transfers. We evaluate the resulting gap equations on the constant-energy contours introduced in Section~\ref{sm:sec:SC}.

\subsection{Magnon Bogoliubov transformation}
The transformation matrices at $\textbf{q}$ and $-\textbf{q}$ contain two representations of the same magnon mode, one as an annihilation operator and the other as a creation operator. Independent diagonalizations can assign phases that are inconsistent with this adjoint relation.

Taking the adjoint of each entry of the magnon spinor and exchanging its two blocks gives the spinor at the opposite momentum,
\begin{equation}
    \boldsymbol{\Psi}_{-\textbf{q}} = \Sigma_x\left(\boldsymbol{\Psi}^\dag_\textbf{q}\right)^T,\qquad
    \Gamma_{-\textbf{q}} = \Sigma_x\left(\Gamma_\textbf{q}^\dag\right)^T,\qquad
    \Sigma_x=\begin{pmatrix}
        0&I_{N_s}
        \\
        I_{N_s}&0
    \end{pmatrix}.
\end{equation}
Applying these relations to $\boldsymbol{\Psi}_\textbf{q}=T(\textbf{q})\Gamma_\textbf{q}$ gives
\begin{equation}
    T(-\textbf{q})=\Sigma_xT^*(\textbf{q})\Sigma_x.
\end{equation}

\subsection{Solution of the gap equations}
\label{sm:sec:Numerical gap}
We obtain the constant-energy contours by using the algorithm described in Appendix C of Ref.~\cite{KL-SE}. For each band \textit{n} and target energy $\xi$, we evaluate $\xi_{\textbf{k}n}-\xi$ on a triangulated momentum grid that extends slightly beyond the magnetic Brillouin zone. When this quantity changes sign across a triangle edge, we locate the contour crossing by linear interpolation. Connecting the crossings within each triangle and joining the resulting segments gives the contours. We then restrict the curves to the magnetic Brillouin zone and resample them at equal intervals in arc length for the contour integrals in the gap equation.

At iteration \textit{p} of the zero-temperature gap equation, we use the current gap $\Delta^{(p)}$ to construct the BdG Hamiltonian and evaluate the right-hand side of \eqref{sm:eq:Zero temp BCS},
\begin{equation}
    \Tilde{\Delta}^{(p+1)} = \mathcal{F}_{T=0}\left[\Delta^{(p)}\right].
\end{equation}
To reduce oscillations between successive iterations, we mix this result with the current gap,
\begin{equation}
    \Delta^{(p+1)}=\left(1-\alpha\right)\Delta^{(p)}+\alpha \Tilde{\Delta}^{(p+1)}.
\end{equation}
We use $\alpha=0.02$ and check convergence using the difference between the current gap and the update before mixing. The iteration stops when
\begin{equation}
    R^{(p)}=\frac{\left|\left|\Tilde{\Delta}^{(p+1)}-\Delta^{(p)}\right|\right|_2}{\left|\left|\Delta^{(p)}\right|\right|_2} <10^{-7},
\end{equation}
where $||\ldots||_2$ is the $2$-norm.

\section{Numerical parameters}
\label{sm:sec:Numerical params}
The magnetic exchange couplings and DMI strength are chosen to give a coplanar ground state with a six-site magnetic unit cell. The easy-axis anisotropy $K_z$ opens a small gap in the magnon spectrum while preserving this spin texture. We choose the on-site electron-spin exchange $J_\mr{sd}$ small compared with the hopping $t$ for a perturbative treatment of the electron-magnon interaction.

Energies are measured in units of the nearest-neighbor hopping \textit{t}, and lengths in units of the lattice constant $a$. We use $t=1$~eV to convert the calculated temperature to Kelvin. Table~\ref{sm:tab:model parameters} lists the parameter values used unless otherwise stated.
\begin{table}[H]
\caption{\label{sm:tab:model parameters}
Physical parameters used in the calculations, unless otherwise specified.}
\begin{ruledtabular}
\begin{tabular}{ccl}
Parameter & Value & Description \\
\hline
$t$ & $1\,\mathrm{eV}$ & Nearest-neighbor hopping \\
$t_2/t$ & $0.25$ & Next-nearest-neighbor hopping \\
$J_{\mathrm{sd}}/t$ & $0.12$ & Electron--spin exchange field strength \\
$\mu/t$ & $-4.3$ & Selected chemical potential \\
$J_x/t$ & $0.01$ & Exchange coupling along $x$ \\
$J_y/t$ & $0.01$ & Exchange coupling along $y$ \\
$D/t$ & $\sqrt{6}/100$ & DMI strength \\
$K_z/t$ & $0.002$ & Easy-axis anisotropy \\
$S$ & $1$ & Localized spin length \\
$a$ & $1$ & Lattice constant \\
\end{tabular}
\end{ruledtabular}
\end{table}

\bibliography{Refs.bib, manual_refs}